\documentclass[11pt,a4paper]{article}
\usepackage{jheppub}
\usepackage{braket} 
\usepackage{multirow}
\usepackage{float}
\usepackage{subcaption}
\usepackage{color}
\usepackage{enumerate}

\begin{document}

\title{Bootstrapping periodic quantum systems}
\author{Zhijian Huang}
\author{and Wenliang Li}
\emailAdd{liwliang3@mail.sysu.edu.cn}
\affiliation{School of Physics, Sun Yat-Sen University, Guangzhou 510275, China}
\abstract{
Periodic structures are ubiquitous in quantum many-body systems and quantum field theories, 
ranging from lattice models, compact spaces, to topological phenomena.
However, 
previous bootstrap studies encountered technical challenges even for one-body periodic problems, 
such as a failure in determining the accurate dispersion relations for Bloch bands. % for a particle in a periodic potential. 
In this work, we develop a new bootstrap procedure to resolve these issues, 
which does not make use of positivity constraints.
We mainly consider a quantum particle in a periodic cosine potential.  
The same procedure also applies to a particle on a circle, 
where the role of the Bloch momentum $k$ is played by the boundary condition or the $\theta$ angle.
We unify the natural set of operators  
and the translation operator 
by a new set of operators $\{e^{inx} e^{iap} p^s\}$. 
To extract the Bloch momentum $k$, 
we further introduce a set of differential equations for $\langle e^{inx} e^{iap} p^s \rangle$ in the translation parameter $a$. 
At some fixed $a$, the boundary conditions can be determined accurately 
by analytic bootstrap techniques and matching conditions. 
After solving the differential equations,  
we impose certain reality conditions to determine the accurate dispersion relations, 
as well as the $k$ dependence of other physical quantities. 
We also investigate the case of noninteger $s$ using the Weyl integral in fractional calculus.}

\maketitle 

\section{Introduction}
The nonperturbative bootstrap has emerged as a powerful approach to study strong coupling physics. 
See \cite{Poland:2018epd,Rychkov:2023wsd} for reviews on the remarkable progress in conformal field theories. 
More recently, there has been considerable interest in extending the bootstrap methods beyond conformal correlators. 
In conformal field theories, one can focus on the macroscopic correlators associated with the infrared fixed point, and carry out the bootstrap studies without knowing the microscopic origins. 
On the other hand, the nonconformal correlators usually depend on the scale and thus the microscopic details. 
\footnote{For gapless systems, one may also study the conformal field theories that emerge in the infrared.}
Accordingly, the self-consistency relations in the bootstrap formulation are associated with a microscopic definition,  
which can be derived in the Lagrangian 
\footnote{The underdetermined nature of the truncated Dyson-Schwinger equations was also discussed recently in \cite{Bender:2022eze,Bender:2023ttu}, where a resolution based on the large $n$ asymptotic behavior was proposed. 
See also \cite{Peng:2024azv} for a perspective from the Lefschetz thimble decomposition and Borel resummation. }
or Hamiltonian formalism. 
See \cite{Lin:2020mme,Han:2020bkb,Hessam:2021byc,Kazakov:2021lel,Lin:2023owt,Li:2024ggr,Khalkhali:2025jzr,Cho:2024kxn,Lin:2024vvg} for the matrix models,  \cite{han2020quantummanybodybootstrap,Aikawa:2021eai,Berenstein:2021loy,Tchoumakov:2021mnh,Berenstein:2021dyf,Bhattacharya:2021btd,Du:2021hfw,Aikawa:2021qbl,Lawrence:2021msm,Blacker:2022szo,Nakayama:2022ahr,Li:2022prn,Khan:2022uyz,Berenstein:2022ygg,Morita:2022zuy,Berenstein:2022unr,Lawrence:2022vsb,Li:2023nip,Guo:2023gfi,Berenstein:2023ppj,Li:2023ewe,John:2023him,Fan:2023tlh,Sword:2024gvv,Li:2024rod,Khan:2024mhc,Gao:2024etm,Robichon:2024apx,Berenstein:2025itw,Aikawa:2025dvt} for the quantum mechanical systems, 
\cite{Anderson:2016rcw,Kazakov:2022xuh,Kazakov:2024ool,Li:2024wrd,Guo:2025fii} for the lattice gauge theories, and \cite{Cho:2022lcj,Nancarrow:2022wdr,Berenstein:2024ebf,Cho:2024owx}  
for the lattice spin models 
that have been considered in the non-conformal bootstrap formulation.

Periodicity is one of the basic structures in physics.  
Before a full-fledged bootstrap investigation of the many-body problems with periodic structures, 
it is a requisite to successfully handle the periodic one-body problems in the bootstrap formulation. 
\footnote{Before studying the properties of excited states, 
the gap between the ground-state and excited-state energies is a fundamental characteristic. 
See for example \cite{Nancarrow:2022wdr} for the related bootstrap studies.   }
However, a direct extension of the quantum mechanical bootstrap method in \cite{Han:2020bkb} 
fails to determine the complete properties of the continuous spectra arising from periodic structures, 
as reported in \cite{Tchoumakov:2021mnh,Berenstein:2021loy,Aikawa:2021eai}. 
Although the physical systems under consideration are different, 
these bootstrap studies found essentially the same difficulty 
in determining the accurate properties of the energy spectra.  
In other words, it is hard to determine the accurate dependence of the energy spectra on
\begin{enumerate}[(i)]
\item
the Bloch momentum for a particle in a periodic potential \cite{Tchoumakov:2021mnh}, 
\item
the boundary condition for a particle on a circle \cite{Berenstein:2021loy}, 
\item
the $\theta$ angle for a charged particle on a circle \cite{Aikawa:2021eai}.  
\end{enumerate}
We refer to \cite{Tchoumakov:2021mnh,Berenstein:2021loy,Aikawa:2021eai} for more details. 
Below, we give a brief overview of the corresponding quantum mechanical systems 
\begin{enumerate}[(i)]

\item {Bloch bands}

The first physical system is a quantum particle in a periodic cosine potential. 
The Hamiltonian reads
\begin{equation}\label{Ham}
    H=p^2+2\alpha\cos x =p^2+\alpha\left( e^{ix}+e^{-ix} \right)\,,
\end{equation}
where $\alpha=1$ denotes the strength of the potential. The period of the potential is $2\pi$. 
In position representation, the momentum operator is given by
\begin{equation}\label{odef}
    p = -i\hbar\frac{d}{dx}\,.
\end{equation}
Below, we set $\hbar=1$.
According to Bloch's theorem, the energy eigenstates are spatially extended states. 
Their wave functions can be viewed as plane waves modulated by periodic functions
\begin{equation}\label{blochth}
    \psi(x)=e^{ikx}\phi_k(x),
\end{equation}
where the period of $\phi_k(x)$ is the same as the period of potential. 
A Bloch momentum $k$ is sometimes called a quasimomentum due to the discrete translational invariance. 
\footnote{To be more precise, $k$ is a wave vector, and the quasimomentum is given by $\hbar k$. }
Under the action of translation operator $T(2\pi)$, we have
\begin{equation}
T(2\pi)\psi(x)=\psi(x+2\pi)=e^{2\pi ik}\psi(x)\,.
\end{equation}
An important consequence of a periodic potential is that 
the energy spectrum is continuous and forms energy bands. 
The energy of a quantum particle varies with the Bloch momentum $k$, 
which is encoded in the dispersion relations. 

\item{Particle on a circle}

The second physical system is a particle on a circle with a cosine potential. 
The Hamiltonian is formally equivalent to \eqref{Ham}. 
The stationary Schr\"odinger equation 
\begin{equation}\label{MSe}
    \frac{d^2}{d x^2}\psi(x)+(E-2 \cos x )\psi(x)=0
\end{equation}
can be written as Mathieu's differential equation after some changes of variables. 
In contrast to the first system, a circle $x+2\pi\equiv x$ is a compact space. 
If the wave function is single valued, we should impose the periodic boundary condition, 
i.e., $\psi(x+2\pi)=\psi(x)$.  
A slight variation is the antiperiodic boundary condition $\psi(x+2\pi)=-\psi(x)$. 
They correspond to \eqref{blochth} with $k=0$ and $k=\pm \frac 1 2$. 
Other real values of $k$ are associated with other phase factors.

\item {Quantum mechanical analog of the $\theta$-term}

The third physical system is a charged particle on a circle, which is similar to the second case.  
However,  the quantum particle here can be affected by magnetic flux.  
If there is a background constant gauge potential, 
then the Aharonov-Bohm effect implies an additional term in the action
\begin{equation}\label{actionqm}
    S(\theta)=\int dt \; \left(\frac{1}{2} \dot{x}^2-V(x)\right) - \frac{\theta}{2\pi} \int dt \; \dot{x},
\end{equation}
where $V(x)$ is a cosine potential
\footnote{There is an additional constant term in the potential of \cite{Aikawa:2021eai}. } and the last term is a quantum mechanical analog of the $\theta$ term. 
As for other topological terms, the $\theta$ term is purely imaginary in the Euclidean action, 
so the Monte Carlo method encounters 
the sign problem or the complex weight problem. 
The Hamiltonian associated with the action \eqref{actionqm} reads
\begin{equation}\label{thetaH}
    H(\theta)=\frac{1}{2}\left(p+\frac{\theta}{2\pi}\right)^2 +V(x),
\end{equation}
where $p$ is defined in \eqref{odef}. 
The presence of a $\theta$ term is equivalent to shifting the momentum $p$ by a constant $\theta/(2\pi)$, 
which is related to the constant gauge potential. 
There is a curious interplay between the boundary condition and the $\theta$ term due to a gauge symmetry of \eqref{thetaH}. 
Suppose that we use the periodic boundary condition $\phi(x+2\pi)=\phi(x)$ to solve the stationary Schr\"odinger equation
\begin{equation}
H(\theta)\phi(x)=E\phi(x)\,.
\end{equation}
By a gauge transformation, we have
\begin{equation}
e^{i\frac {\lambda}{2\pi}x}H(\theta)e^{-i\frac {\lambda}{2\pi}x}\left[e^{i\frac {\lambda}{2\pi}x}\phi(x)\right]=
H(\theta-\lambda)\left[e^{i\frac {\lambda}{2\pi}x}\phi(x)\right]=E \left[e^{i\frac {\lambda}{2\pi}x}\phi(x)\right]\,.
\end{equation}
If we set $\lambda=\theta$, then the Hamiltonian becomes $H(0)=p^2/2+V(x)$.  
The $\theta$ term effects are encoded in the boundary condition
$e^{i\frac {\theta}{2\pi}(x+2\pi)}\phi(x+2\pi)=e^{i {\theta}} [e^{i\frac {\theta}{2\pi}x}\phi(x)]$, 
so the dependence on the $\theta$ angle is similar to that on the Bloch momentum $k$. 
Another comment is that the gauge invariant version of $\langle{p}\rangle$ is the velocity  
$\langle {\dot x}\rangle\equiv\langle {p}\rangle+\theta/(2\pi)$.
\end{enumerate}

As the bootstrap formulation does not make use of the wave functions explicitly, 
it is less straightforward to identify the Bloch momentum, the boundary conditions, and the $\theta$ angle 
in the above physical systems. 
Since the Bloch momentum $k$ can be readily translated into 
the boundary condition or the $\theta$ angle for a particle on a circle,  
we focus on the first system and do not repeat the discussion for the latter two systems. 
Our main goal is to develop a bootstrap procedure that can capture 
the accurate dispersion relation of a quantum particle in a periodic potential, 
whose explicit Hamiltonian is given in \eqref{Ham}. 

Let us briefly summarize the previous bootstrap results in \cite{Tchoumakov:2021mnh}. 
The authors considered the set of operators 
\begin{equation}
\{e^{inx}p^s\}\,,\quad
n=0,\pm 1,\pm 2, \dots\,,\quad s=0,1,2,\dots,
\label{natural-set}
\end{equation}
and derive the recursion relations for their expectation values. 
After fixing the normalization and solving the recursion relations, there remain three independent parameters 
$(E,\, \langle{e^{ix}}\rangle, \langle{p}\rangle)$. 
The authors in \cite{Tchoumakov:2021mnh} considered:
\begin{equation}\label{Oeixp}
    \mathcal{O}=\sum^K_{n=0} \sum^L_{s=0} a_{ns} p^{s} e^{inx}, 
\end{equation}
where $\{a_{ns}\}$ are arbitrary complex coefficients and $(K, L)$ are positive integer numbers. 
They set $L=1$ and 
used the positivity constraint $\langle{\mathcal{O}^\dagger\mathcal{O}} \rangle\geq 0$ 
to derive the physical regions in the two-dimensional parameter space 
$(E, \langle{e^{ix}}\rangle)$ by setting $\langle{p}\rangle=0$. 
As the truncation parameter $K$ increases, the predictions become more precise   
and the bounds converge to the accurate results from the standard diagonalization method. 
However, the dispersion relation concerns the energy dependence on 
the Bloch momentum $k$, which is related to $e^{2\pi i p}$, i.e., an exponential of the momentum operator $p$. 
Na\"ively, one can express $e^{2\pi i p}$ in terms of $\{p^s\}$ using the Taylor series
\begin{equation}\label{blochk}
    e^{2\pi i k}=\langle{e^{2\pi i p}}\rangle=\sum_{s=0}^{\infty}\frac{(2\pi i)^s}{s!}\langle{p^s}\rangle\,.
\end{equation}
However, $p$ is not a bounded operator, 
so the high moments of $p$ have large errors and the result from \eqref{blochk} is unstable.  
To resolve this issue, 
the authors in \cite{Tchoumakov:2021mnh} derived $\langle{e^{2\pi i p}}\rangle$ from the probability distributions 
that can reproduce a finite number of low moments $\langle{p^s}\rangle$. 
We refer to \cite{Tchoumakov:2021mnh} for more details on the probability distributions. 
Unfortunately, the resulting dispersion relations still exhibit significant deviations from the diagonalization results.
Some parallel bootstrap issues for the other two systems were discussed in \cite{Berenstein:2021loy, Aikawa:2021eai}. 
Similarly, they noticed an obstacle in extracting the boundary condition or the $\theta$ angle, 
and thus the accurate properties of the energy spectra. 

In \cite{Tchoumakov:2021mnh,Berenstein:2021loy, Aikawa:2021eai}, 
it is natural to study the set of operators in \eqref{natural-set},  
as the Hamiltonian \eqref{Ham} is composed of a monomial in $p$ and exponentials in $x$. 
However, the momentum exponential operator $e^{2\pi ip}$ does not belong to a finite span of this natural set of operators. 
\footnote{For the Kronig-Penney model, 
the infinite sums can be evaluated using the Riemann zeta function \cite{Blacker:2022szo}, 
leading to an analytic derivation of the dispersion relation. 
It is not clear to us how to apply this approach to the cases without simple analytic solutions. }
This motivates us to expand the set of operators under consideration.  
Although the periodic system \eqref{Ham} has only discrete translation symmetry, 
we consider the continuous translation operator
\begin{equation}
T(a)\equiv e^{iap}\,.
\end{equation} 
The Bloch momentum is associated with the cases in which $a$ is an integer multiple of the period, 
such as the case in \eqref{blochk}. 
Therefore, we consider a larger set of operators
\begin{equation}
\label{larger-operator-set}
\{e^{inx} e^{iap} p^s\}\,,\quad
n=0,\pm 1,\pm 2, \dots\,,\quad s=0,1,2,\dots, \quad
a\in [0,2\pi]\,.
\end{equation}
We can restrict the range of $a$ to $[0,2\pi]$ because $T(a+2\pi)\psi(x)=e^{2\pi ik}T(a)\psi(x)$. 

The reason for considering a continuous range of $a$ is as follows. 
The Bloch momentum is related to the additional phase from a $2\pi$ translation, 
such as 
\begin{equation}
k=\frac 1 {2\pi i}\ln \frac{\langle{T(2\pi)}\rangle}{\langle{T(0)}\rangle}\,.
\end{equation}
Since the Hamiltonian \eqref{Ham} does not involve any exponential term in $p$,  
we cannot use the recursion relations to constrain the $a$ dependence. 
In other words, the observables with different $e^{iap}$ ``backgrounds'' decouple. 
As a finite change in $a$ is not available, we switch to an infinitesimal change in $a$,  
i.e., the derivative with respect to $a$.  
It is simple to deduce a set of first-order differential equations in $a$ from operator equations
\begin{equation}
\label{diff-eq}
\frac {\partial}{\partial a}\langle {e^{inx} e^{iap} p^s}\rangle=i\langle {e^{inx} e^{iap} p^{s+1}}\rangle\,,
\end{equation}
where the operators on the right-hand side still belong to \eqref{larger-operator-set}. 
To solve these first-order differential equations, 
we need to impose some boundary conditions at a specific $a$. 
For a given $a$, we can use the large $n$ expansion and matching conditions \cite{Li:2023ewe,Li:2024rod} 
to derive analytic solutions to the recursion relations, which are approximate but highly accurate. 
We still need to be careful about the unphysical divergences in the $a\rightarrow0$ limit of the differential equations, 
which can be removed by imposing regularity of the small $a$ expansion. 
\footnote{This is similar to the regularity conditions for the small coupling expansion in \cite{Guo:2023gfi}. } 
The natural choices for the location of the boundary conditions are $a=0$ and $a=\pi$. 
The first choice reduces to the smaller set of operators in 
\cite{Tchoumakov:2021mnh,Berenstein:2021loy, Aikawa:2021eai}. 
The second choice corresponds to the case of half-period translation, 
which is special due to a discrete symmetry. 
Using the boundary conditions at a specific $a$, 
we can solve the differential equations \eqref{diff-eq} numerically. 
Since the expectation values are periodic in $a$ up to a phase factor, 
we can also use truncated Fourier series to approximate their dependence on $a$. 
In the latter approach, 
the analytic dependence on $a$ allows us to further investigate $\langle{p^s}\rangle$ with noninteger power $s$, 
which is related to the Weyl integral in fractional calculus. 

Let us briefly explain the last ingredient: reality conditions. 
At a given energy $E$, 
the solutions to the differential equations do not determine the physical values of $k$. 
Instead, we only obtain the relationship between $k$ and $\langle{p}\rangle$. 
To extract the Bloch momenta, 
we further require that $(E, k, \langle{p}\rangle)$ should be real numbers. 
For instance, if we find that $k$ and $\langle{p}\rangle$ cannot be real at the same time, 
then the energy $E$ is in a band gap. 
Using the reality conditions, 
we can derive the accurate range of the energy bands and their dispersion relations, 
together with the $k$ dependence of other expectation values. 

The paper is organized as follows. 
In Sec. \ref{secBA}, we provide more details about the bootstrap formulation, 
such as the recursion relations and the differential equations. 
Then we solve them analytically using the large $n$ expansion and small $a$ expansion. 
In Sec. \ref{secBRa}, we study the recursion relations nonperturbatively at special $a$ 
using the analytic bootstrap techniques and matching conditions. 
In Sec. \ref{secBR}, we solve the differential equations and extract the physical Bloch momenta $k$ from the reality conditions, 
which resolve the technical challenges in \cite{Tchoumakov:2021mnh, Berenstein:2021loy, Aikawa:2021eai}.
In Sec. \ref{nonints}, we study $(ip)^s$ with noninteger $s$ using the Weyl integral in fractional calculus. 
In the end, we summarize our results and discuss some future directions in Sec. \ref{conclusion}.

\section{Analytic analysis of self-consistency conditions}\label{secBA}
In this work, 
we consider a new set of operators $\{e^{inx} e^{iap} p^s\}$ 
with $n=0,\pm 1,\pm 2, \dots$,  $s=0,1,2,\dots$, and $a\in [0,2\pi]$. 
As discussed above, the operators considered in the previous studies \cite{Tchoumakov:2021mnh,Berenstein:2021loy, Aikawa:2021eai} are the special cases with $a=0$. 
To simplify the notation, we denote the expectation values as
\begin{equation}
f_{n,a,s}\equiv\langle{e^{inx} e^{iap} p^s}\rangle\,,
\end{equation}
which are associated with an energy eigenstate labeled by $E$. 
Furthermore, we omit the third subscript if $s=0$
\begin{equation}
f_{n,a}\equiv \langle{e^{inx} e^{iap}}\rangle\,.
\end{equation}
In Sec. \ref{Rec}, we derive some recursion relations for $f_{n,a}$ in $n$, % and those for $f_{n,a}$ in $n$, 
which are based on the definition of the Hamiltonian \eqref{Ham}. 
In Sec. \ref{TDE}, we use operator equations to deduce some first-order differential equations in $a$, 
which connect the expectation values $f_{n,a}$ at different $a$. 
In some special limits, these self-consistency conditions are simplified and can be studied analytically. 
We discuss the large $n$ expansion in Sec. \ref{Lnex} and the small $a$ expansion in Sec. \ref{sae}. 

\subsection{Recursion relations for $\langle{e^{inx} e^{iap}}\rangle$ in $n$}
\label{Rec}
For an energy eigenstate with real $E$, 
the expectation values satisfy the following self-consistency conditions
\begin{equation}\label{cons1}
\langle{[H,\mathcal{O}]}\rangle=0\,,
\end{equation}
\begin{equation}\label{cons2}
\langle{\mathcal{O}H}\rangle=E\langle{\mathcal{O}}\rangle\,.
\end{equation}
If the Hamiltonian is not self-adjoint, i.e., $H\neq H^\dagger$, 
an anomaly term $\mathcal{A}_\mathcal{O}\equiv\langle{(H^\dagger-H)\mathcal{O}}\rangle$ may appear in \eqref{cons1}, 
as emphasized in \cite{Berenstein:2022ygg} (see also \cite{Esteve:2002nt,Juric:2021psr}). 
For the extended Bloch states, we restrict the integration domain of the inner product to a finite interval, 
so we need to be careful about the anomaly term. 
For the concrete Hamiltonian \eqref{Ham}, the explicit expression of the anomaly term is
\begin{equation}\label{anom}
\mathcal{A}_\mathcal{O}\propto 
\left. \left[\psi^*\frac{d( \mathcal{O} \psi)}{dx}
-\frac{d\psi^*}{dx} \mathcal{O} \psi
\right]\right|_{x_1}^{x_2}\,,
\end{equation}
where $\psi(x)$ is the wave function and the integration domain is $[x_1, x_2]$. 
For the operators in \eqref{larger-operator-set}, 
we choose $x_1=0$ and $x_2=2\pi$ to avoid the anomaly term
\footnote{For $\mathcal O=e^{inx} e^{iap} p^s$ with noninteger $n$, we should enlarge the integration domain. For a rational number $n=n_1/n_2$, we can choose $x_2=2\pi n_2$. 
For example, the integration domain for $n=2/3$ becomes $[0, 6\pi]$. 
If $n$ is an irrational number, we can consider a rational approximation for $n$ and remove the anomaly term.  
Then we take the limit where the error of the rational approximation vanishes.
In the diagonalization method, we choose the integration domain as described above.} 
\begin{equation}
\mathcal{A}_\mathcal{O}=0\,.
\end{equation}

As we consider the exponentials of both $x$ and $p$, 
it is useful to consider the Baker-Campbell-Hausdorff formula
\begin{equation}
\label{BCH}
    e^{A}e^{B}=e^{A+B+\frac{1}{2!}[A,B]+ \dots}\,,
\end{equation}
where we have omitted the terms irrelevant to our discussion. 
For $A, B\in(inx, iap)$, we can deduce $e^{A}e^{B}=e^{[A,B]}e^{B}e^{A}$ from \eqref{BCH}, 
so the commutation relation between $e^{inx}$ and $e^{iap}$ reads
\begin{equation}\label{hauscm}
    [e^{inx},e^{iap}]=(1-e^{ina})e^{inx}e^{iap}.
\end{equation}
Some basic operator identities are also useful
\begin{equation}
    [A,BC]=[A,B]C+B[A,C],
\quad
    [p,e^{inx}]=ne^{inx}\,.
\end{equation}
For $\mathcal{O}=e^{inx}e^{iap}$, the self-consistency conditions \eqref{cons1}, \eqref{cons2} 
associated with  \eqref{Ham}  are
\begin{equation}\label{rela1}
    n^2 f_{n,a}+2n f_{n,a,1}+\left(1- e^{ia}\right) f_{n+1,a}+\left(1- e^{-ia} \right) f_{n-1,a}=0\,,
\end{equation}
and
\begin{equation}\label{rela2}
f_{n,a,2} + e^{ia}f_{n+1,a}+ e^{-ia}f_{n-1,a}=E f_{n,a}\,.
\end{equation}
According to \eqref{diff-eq}, the derivative of \eqref{rela1} with respect to $a$ gives
\begin{align}\label{rela3}
&n^2 f_{n,a,1} +  2nf_{n,a,2}
+\left(1- e^{ia} \right) f_{n+1,a,1}+\left(1- e^{-ia}\right) f_{n-1,a,1}- e^{ia}f_{n+1,a} + e^{-ia}f_{n-1,a}=0\,.
\end{align}
It is straightforward to generate recursion relations for higher $s$ by taking a higher-order $a$ derivative. 
If $e^{ia}=1$, the coefficients of some terms in \eqref{rela1} and \eqref{rela3} vanish.  
Let us first consider the special case $e^{ia}=1$ and then discuss the generic case $e^{ia}\neq 1$. 

For $e^{ia}=1$, there are two possibilities in the range $a\in [0,2\pi]$: $a=0$ and $a=2\pi$. 
For $a=0$, we can use \eqref{rela1}, \eqref{rela2} and \eqref{rela3} to derive 
a recursion relation for the $s=0$ terms
\begin{equation}\label{a0rela}
2(2 n+1)f_{n+1,0}+n(n^2 - 4E) f_{n,0}+    2(2 n-1) f_{n-1,0}=0,
\end{equation}
which is invariant under 
$n \to -n$ and $f_{n^\prime,0} \to f_{-n^\prime,0}$. 
We choose the independent set of free parameters as
\begin{equation}
    (E,\, f_{0,0}=\langle{1}\rangle,\, f_{1,0}=\langle{e^{ix}}\rangle).
\end{equation}
It is natural to impose the normalization condition
\begin{equation}
\label{normalization}
    f_{0,0}=1. 
\end{equation}
Some explicit solutions at small $|n|$ are
\begin{equation}\label{O0rela}
   	f_{-1,0}=f_{1,0}\,,\quad f_{\pm 2,0}=\frac{1}{6} (4 E-1) f_{1,0}-\frac{1}{3}\,,\quad
        f_{\pm 3,0}=\frac{1}{15} (2E+1) (4E-7)f_{1,0}-\frac{4}{15} (E-1)\,.
\end{equation}
As $\langle{e^{inx}}\rangle=\langle{e^{-inx}}\rangle$, we can also interpret $f_{n,0}$ as $\langle{\cos (n x)}\rangle$, which is used in \cite{Tchoumakov:2021mnh}. 
In the explicit solutions for $f_{n,0}$, the degree of $E$ grows with $|n|$, but $f_{n,0}$ is at most linear in $f_{1,0}$. This feature is also shared by the cases with $a\neq 0$. 
As the $a=0$ recursion relation is simpler than the generic case, 
the corresponding solutions also exhibit simpler structures. 
We use the $a=0$ solutions as boundary conditions for the differential equations in $a$, 
which is discussed in Sec. \ref{M1}. 

For $a=2\pi$, we have
\begin{equation}\label{recur2pi}
   2 (2 n+1) f_{n+1,2\pi}+n(n^2-4E) f_{n,2\pi}+2(2n-1)f_{n-1,2\pi}=0\,.
\end{equation}
Bloch's theorem indicates $f_{n,2\pi}=e^{2\pi i k}f_{n,0}$, 
so the recursion relation for $a=2\pi$ is equivalent to that for $a=0$ in \eqref{a0rela}. 
The main difference is that we cannot simply set the $n=0$ case of $f_{n,2\pi}$ to $1$. 
In fact,  the Bloch momentum $k$ is precisely encoded in $f_{0,2\pi}=e^{2\pi i k}$ 
if we use the normalization condition \eqref{normalization}. 
The recursion relations associated with the Hamiltonian do not provide a connection between the case of $a=0$ and $a=2\pi$, so we cannot use them to determine the Bloch momentum, as noted in \cite{Tchoumakov:2021mnh,Berenstein:2021loy,Aikawa:2021eai}. 
Before studying the continuous dependence on $a$, 
let us discuss the case of generic $a$. 

For $e^{ia}\neq 1$, we can also derive a recursion relation for the $s=0$ terms from \eqref{rela1}, \eqref{rela2}, and \eqref{rela3},
\begin{align}\label{recrelation}
    & n\left[ (n^2-1)(n^2-4E)+4(1-\cos a)\right]f_{n,a}  \nonumber\\
    +& (n^2-1)\left[(2n+1)(1+e^{ia})f_{n+1,a}+(2n-1)(1+e^{-ia})f_{n-1,a} \right]
    \nonumber\\
    +&(n-1)(1-e^{ia})^2 f_{n+2,a}+(n+1)(1-e^{-ia})^2 f_{n-2,a} =0\,,
\end{align}
which is invariant under $n\rightarrow -n$, $a\rightarrow -a$, $f_{n',a'}\rightarrow f_{-n',-a'}$. 
We can choose the independent set of free parameters as
\begin{equation}\label{seta}
    (E,\,f_{0,a},\,f_{1,a},\,f_{2,a},\, f_{3,a},\,f_{-3,a}).
\end{equation}
As \eqref{recrelation} is a fourth-order difference equation, one may think that four $f_{n,a}$'s are sufficient, but the recursion relation \eqref{recrelation} for $n=-1, +1$ are not independent,
\footnote{Furthermore, the coefficients of $f_{\pm 3,a}$ vanish, so the initial conditions with four consecutive $f$'s only work in one direction for sufficiently large $|n|$.  } so we have one more free parameter. 
The explicit solutions are more involved due to the presence of $(e^{ia}, e^{-ia})$. 
Some explicit solutions for small $|n|$ are
\begin{equation}
f_{-1,a}=e^{ia}f_{1,a}\,,\quad f_{-2,a}=e^{2ia}f_{2,a}\,,
\end{equation}
\begin{equation}\label{f4a}
f_{\pm 4,a}=-3e^{\mp 2ia}f_{0,a}-\frac {9(1+e^{\mp ia})}{(1-e^{\pm ia})^2}f_{\pm 1,a}+
\frac {8(3E-4+\cos a)}{(1-e^{\pm ia})^2}f_{\pm 2,a}
-\frac {15(1+e^{\pm ia})}{(1-e^{\pm ia})^2}f_{\pm 3,a}\,.
\end{equation}
If we set $e^{ia}=1$, then the coefficients of $(f_{n-2,a}, f_{n+2,a})$ in \eqref{recrelation} vanish,  
and the explicit equation reduces to \eqref{a0rela} or \eqref{recur2pi} with an additional factor $(n^2-1)$. 
\footnote{The additional factor $(n+1)(n-1)$ is not needed 
in the derivation of \eqref{a0rela} or \eqref{recur2pi} 
because the coefficients of $f_{n\pm1,a,1}$ in \eqref{rela3} vanish at $e^{ia}=1$. }
Then one can notice that the case of $a=\pi$ is also special in that the coefficients of $f_{n\pm 1,a}$ vanish. 

For $a=\pi$, the recursion relation \eqref{recrelation} becomes
\begin{equation}\label{recpi}
     4 (n-1) f_{n+2,\pi }+ n \left((n^2-1)(n^2-4E)+8 \right) f_{n,\pi }+4 (n+1) f_{n-2,\pi }=0\,,
\end{equation}
which is invariant under $n\rightarrow -n$, $f_{n',\pi}\rightarrow f_{-n',\pi}$.
Note that the odd $n$ cases and the even $n$ cases form two independent sectors. 
Some explicit solutions are
\begin{align}
    	f_{-1,\pi}&=-f_{1,\pi}\,,\quad f_{-2,\pi}=f_{2,\pi}\,,\\
        f_{\pm 4,\pi} &= -3 f_{0,\pi} + 2 (3 E-5) f_{2,\pi},
        \quad f_{\pm 5,\pi} = -2 f_{\pm 1,\pi} + 6 (2 E-5) f_{\pm 3,\pi},\\
        f_{\pm 6,\pi} &= -4(15 E -62) f_{0,\pi} + 3\left(40 E^2-232 E+275\right) f_{2,\pi},\\
        f_{\pm 7,\pi} &= -20(3E -19) f_{\pm 1,\pi} + \frac 3 2 (240 E^2-2120 E+3799) f_{\pm 3,\pi}\,. 
\end{align}
As \eqref{recpi} shares some similarities with \eqref{a0rela}, 
it also natural to use the $a=\pi$ solutions as boundary conditions 
for the differential equations in $a$, 
which is discussed in Sec. \ref{M2}. 

\subsection{Differential equations for $\langle{e^{inx} e^{iap}}\rangle$ in $a$}\label{TDE}
Above, we derived the recursion relation for $f_{n,a}$ in $n$ at various $a$.  
Let us also deduce the differential equations that control the $a$ dependence. 
According to \eqref{diff-eq}, the first-order differential equations for $\langle{e^{inx} e^{iap}}\rangle$ are
\begin{equation}\label{df}
    \frac{\partial}{\partial a}f_{n,a}=i f_{n,a,1}\,,
\end{equation}
which are linear equations 
because we can express $f_{n,a,1}$ in terms of a linear combination of $f_{n,a}$ using the recursion relations in Sec. \ref{Rec}. 
For example, we can use \eqref{rela1} to express the right-hand side of \eqref{df} in terms of $(f_{n-1,a}, f_{n,a}, f_{n+1,a})$
\begin{equation}\label{dif}
    \frac{\partial}{\partial a}f_{n,a}=-\frac{i}{2n}\left[\left(1- e^{-ia} \right) f_{n-1,a} +n^2 f_{n,a}+ \left(1- e^{ia} \right) f_{n+1,a}\right]\,,
\end{equation}
which is valid for $n\neq 0$. 
As our goal is to determine the relation between $f_{0,0}$ and $f_{0,2\pi}$, 
we can restrict to the non-negative cases, i.e., $n\geq 0$. 
Then the negative $n$ sector is irrelevant to our discussion, 
so we do not consider the independent parameter $f_{-3,a}$ in \eqref{seta}. 
We choose the set of independent expectation values as
\begin{equation}
\mathcal{F} \equiv(f_{0,a}\quad f_{1,a}\quad f_{2,a}\quad f_{3,a})^\text{T}\,.
\end{equation}
Their differential equations can be written in a matrix form
\begin{equation}\label{pde0}
    \frac{\partial}{\partial a}\mathcal{F}=i\mathcal{M}\mathcal{F}\,. 
\end{equation}
The explicit matrix elements of $\mathcal{M}$ are
\begin{equation}\label{M0123}
\mathcal{M}=
\begin{pmatrix}
\frac{ 1+e^{-i a}}{2 \left(1-e^{-i a}\right)} 
& -\frac{e^{-i a}+4(2 E-1)+e^{i a}}{4 \left(1-e^{-i a}\right)}
& \frac{3  \left(1+e^{i a}\right)}{2 \left(1-e^{-i a}\right)} 
& -\frac{1-e^{i a}}{4e^{-i a} } \\
-\frac{1-e^{-i a}}{2} 
& -\frac{1}{2} 
& -\frac{1-e^{i a}}{2} 
& 0 \\
0 
& -\frac{1-e^{-i a}}{4} 
& -1 
& -\frac{1-e^{i a}}{4} \\
-\frac{1-e^{-i a}}{2e^{i a}}  
& \frac{3 \left(1+e^{-i a}\right)}{2(1- e^{i a})} 
& -\frac{ e^{-i a}+2 (4E-5)+e^{i a}}{2 \left(1-e^{i a}\right)} 
& \frac{ 1+4 e^{i a}}{1-e^{i a}} \\
\end{pmatrix}
\,.
\end{equation}
The second and third rows can be extracted from \eqref{dif}. 
To derive the first and fourth rows, we need to use \eqref{rela1}, \eqref{rela2} and \eqref{rela3}, 
so the energy $E$ appears in the coefficients of the differential equations \eqref{pde0}. 

For consistency, the differential equations for $n>3$ should be redundant. 
We examine some concrete cases 
and verify that they indeed reduce to \eqref{pde0} if the recursion relations are satisfied.

\subsection{Large $n$ expansion}\label{Lnex}
To solve the differential equations \eqref{pde0}, 
we need to impose some boundary conditions at some fixed $a$. 
As the recursion relations in Sec. \ref{Rec} are underdetermined, 
we need to introduce additional constraints. 
One approach is to implement the positivity constraints \cite{Tchoumakov:2021mnh,Berenstein:2021loy,Aikawa:2021eai}, 
but the positivity bounds provide only numerical results. 
Below, we use an analytic method, i.e., the large $n$ expansion.  
Together with the matching conditions in Sec. \ref{secBRa}, 
we can derive highly accurate solutions for $f_{n,a}$, which are given by rational functions in $E$. 

As $n$ increases, we expect that $f_{n,a}$ decays rapidly 
due to the highly oscillatory term $e^{inx}$ in the finite-interval integral
\begin{equation}
    |f_{n+1,a}| \ll |f_{n,a}|\quad (n\rightarrow \infty).
\end{equation}
Accordingly, the leading terms in the recursion relation \eqref{recrelation} are
\begin{equation}\label{largenrec}
     -n^5 f_{n,a} \sim n^2(2n-1)(1+e^{-ia}) f_{n-1,a} + (n+1)(1-e^{-ia})^2 f_{n-2,a}
     \quad (n\rightarrow \infty)\,,
\end{equation}
where we take into account some subleading terms to simplify the discussion. 
It seems that \eqref{largenrec} is still a bit complicated, 
so let us examine the two special cases mentioned in Sec. \ref{Rec}
\begin{align}
a=0:&\quad -n^5 f_{n,0}\sim 2n^2(2n-1) f_{n-1,0}\,, \\
a=\pi:&\quad
    -n^5 f_{n,\pi} \sim 4(n+1) f_{n-2,\pi}\,. 
\end{align}
Since there is only one term on the right-hand side, 
it is not hard to find their solutions 
\begin{align}\label{lnef0pi}
    f_{n,0} \sim 
    &\frac{ 2\pi\Gamma\left(n+\frac 1 2\right)}{\Gamma(n+1)^3} (-4)^{n}\lambda_0
    \sim e^{2 n} n^{-2 n} n^{-\frac 3 2}  (-4)^{n} \lambda_0\,, \\    
    f_{n,\pi} \sim &\frac{\sqrt 2\, \pi^2\Gamma \left(\frac{n+3}{2}\right)}{ \Gamma \left(\frac{n+2}{2}\right)^5}  \left(\frac i 2\right)^n\left(\lambda_1+(-1)^n\lambda_2\right)
    \sim e^{2 n} n^{-2 n} n^{-\frac 3 2} \left[(2i)^{n} \lambda_1+(-2i)^{n} \lambda_2\right]\,, 
\end{align}
where $\lambda_0, \lambda_1,\lambda_2$ are constants. 
Based on these large $n$ asymptotic behaviors,  
we can guess a general expression
\begin{equation}\label{lnef}
    f_{n,a}\sim (-1)^ne^{2 n} n^{-2 n} n^{-\frac 3 2} \left[\left(1+e^{\frac{-ia}{2}}\right)^{2n} \xi_{+,a}+\left(1-e^{\frac{-ia}{2}}\right)^{2n} \xi_{-,a}\right]\,,
\end{equation}
where $\lambda_0,\;\lambda_1,\;\lambda_2$ are replaced by the $a$-dependent functions $\xi_{+,a}=\xi_+(a)$ and $\xi_{-,a}=\xi_-(a)$. 
The fact that the second branch vanishes for $a=0$
also explains the reduced number of free parameters at $a=0$. 

It turns out that \eqref{lnef} is the correct leading behavior at large $n$ for generic $a$. 
Then we can systematically deduce the subleading terms from the complete recursion relation \eqref{recrelation}
\begin{equation}\label{hoc}
        f_{n,a} \sim   (-1)^ne^{2 n} n^{-2 n} n^{-\frac 3 2} 
        \sum_{\sigma=\pm 1}    \xi_{\sigma,a} \left(1+\sigma e^{\frac{-ia}{2}}\right)^{2n} 
      \left( 1+ \sum_{j=1}^{N}c_{\sigma,a}^{(j)}\,n^{-j} \right) \,,
\end{equation}
where $N$ is the truncation order of the $1/n$ series. 
The general form of the coefficients is
\begin{equation} \label{cpmja}
c_{\pm,a}^{(j)}=\sum_{l=0}^{j} \tilde c^{(j)}_{\pm,l}(E)\cos\frac {la}{2}\,,
\end{equation}
where $\tilde c^{(j)}_{\pm,l}(E)$ are degree-$j$ polynomials in $E$. 
Some explicit solutions for $c^{(j)}_{\pm,a}$ are
\begin{align}\label{hocc1}
        c_{\pm,a}^{(1)}=&\,-\frac{5}{12}-2 E\pm\left(\frac{1}{8}-2 E\right) \cos \frac{a}{2},\\
        c_{\pm,a}^{(2)}=&\,3 E^2+\frac{53 E}{24}+\frac{281}{2304}\pm\left(4 E^2+\frac{19 E}{12}-\frac{11}{96}\right) \cos \frac{a}{2}+\left(E^2-\frac{5E}{8}+\frac{9}{256}\right) \cos a\,.
\end{align}
Based on the explicit solutions, we also notice that $c_{\pm,a}^{(j)}=c_{\mp,a+2\pi}^{(j)}$. 
To show this identity, let us replace $a$ with $a+2\pi$
\begin{equation}\label{hoc2pi}
        f_{n,a+2\pi} \sim   (-1)^n e^{2 n} n^{-2 n} n^{-\frac 3 2} 
        \sum_{\sigma=\pm 1} \xi_{\sigma,a+2\pi} \left(1-\sigma e^{\frac{-ia}{2}}\right)^{2n}
         \left( 1+ \sum_{j=1}^{N}c_{\sigma,a+2\pi}^{(j)}\,n^{-j} \right) \,.
\end{equation}
Bloch's theorem implies 
$
f_{n,a+2\pi}=e^{2\pi i k}f_{n,a}
$, so we have
\begin{equation}\label{xi2pi}
    \xi_{\pm,a+2\pi} = e^{2\pi i k}\xi_{\mp,a}\,,
\quad 
    c_{\pm,a+2\pi}^{(j)}=c_{\mp,a}^{(j)}\,.
\end{equation}
Furthermore, we can use the differential equations to determine the $a$ dependence of the prefactors $\xi_\pm(a)$.
If we substitute \eqref{hoc2pi} into \eqref{dif}, 
the large $n$ expansion gives rise to two first-order ordinary differential equations in $a$
\begin{equation}\label{derivexi}
    \xi'_{+,a} +\frac{ \tan(a/4)}{4}\xi_{+,a}=0\,,\quad
    \xi'_{-,a}-\frac{ \cot(a/4)}{4}\xi_{-,a}=0\,.
\end{equation}
The solutions are
\begin{equation}\label{bpm}
    \xi_{+, a}= b_+\cos \frac{a}{4},\quad \xi_{-,a}= b_-\sin \frac{a}{4},\quad a\in [0,2\pi]\,,
\end{equation}
where $b_+$ and $b_-$ are $a$-independent constants. 
As $f_{n,0}$ is real for integer $n$,  
the constant $b_+$ should also be a real number.
According to \eqref{xi2pi}, we have $\xi_{-,2\pi}=e^{2\pi i k}\xi_{+,0}$, so the relation between $b_+$ and $b_-$ is
\begin{equation}\label{bkb}
    b_-=e^{2\pi i k } b_+\,,
\end{equation}
where we assumed that \eqref{bpm} is valid for $a\in [0,2\pi]$. 
We verify that these results are consistent with the diagonalization results 
and satisfy the recursion relations \eqref{rela1}, \eqref{rela2}, and \eqref{rela3}. 
There remain three parameters in the large $n$ solution for general $a$
\begin{equation}\label{generic-a-large-n}
(E,\quad k,\quad b_+)\,.
\end{equation} 
If the $1/n$ series provide accurate approximations at relatively large $n$ around $a=0, 2\pi$, 
then we may use the matching conditions and 
the normalization condition \eqref{normalization}   
to determine $b_+$ 
and the additional free parameters at small $n$. 
However, there are two subtle issues around $a=0,2\pi$:
\begin{enumerate}[(i)]
    \item 
    According to the generic $a$ recursion relation \eqref{recrelation}, 
    the nonperturbative solutions for 
    $f_{n,a}$ contain $(1-e^{ia})^{-2(|n|-3)}$ terms,  
    which are divergent in the $e^{ia}\rightarrow 1$ limit. 
    See \eqref{f4a} for the explicit example of $f_{4,a}$.  
     \item 
    According to the $a$ dependence of the prefactors \eqref{bpm},  
    one of them vanishes in the $a\rightarrow 0, 2\pi$ limits. 
    Then we cannot use the relative phase of the prefactors $\xi_{+,a}$ and $\xi_{-,a}$ 
    to encode the Bloch momentum $k$.   
\end{enumerate}
A natural resolution is to consider the large $n$ expansion and matching conditions in the region around $a=\pi$, 
which is far from the singular limits $a=0, 2\pi$. 
Then we use the differential equations 
to build the connection between $f_{n,a}$ at $a=\pi$ and $a=0, 2\pi$.  
On the other hand, we can also start from the boundaries at  $a=0, 2\pi$. 
Although the $a\rightarrow 0,2\pi$ limits seem problematic, 
the recursion relations for $a=0,2\pi$ are well defined. 
There are no divergent terms $(1-e^{ia})^{-2(|n|-3)}$ 
due to the simple structure of the recursion relations at $a=0,2\pi$. 
The fact that the $a=0$ recursion relation can be studied 
by the large $n$ expansion and matching conditions was briefly noted in \cite{Li:2024rod}. 
To resolve the second issue, we need to consider the $a\neq 0, 2\pi$ region, 
and extract the Bloch momentum $k$ from the $a$ dependence. 

In both approaches, we still need to deal with the unphysical divergences in the differential equations near $a=0,2\pi$. 
Below, we explain how to resolve this issue by the small $a$ expansion. 

\subsection{Small $a$ expansion}\label{sae}
In Sec. \ref{Rec}, we mentioned that the natural boundary conditions are at $a=0$ and $a=\pi$. 
In fact, the recursion relation at $a=0$ is particularly simple.  
For instance, the number of independent free parameters is reduced. 
On the other hand, the simplification at $a=0$ is closely related to the divergent behavior of the differential equations near $a=0$. 
In the coefficient matrix \eqref{M0123}, the divergences manifest as the $(1-e^{\pm i a})^{-1}$ poles. 

However, $\{f_{n,0}\}$ should be finite, so we impose that the $a\rightarrow 0$ limit is regular. 
In other words,  $f_{n,a}$ admits a power series expansion in small $a$,
\begin{equation}\label{smallao} 
    f_{n,a}\sim\sum_{j=0}^{A} f_{n,0}^{(j)}\, a^j\, ,
\end{equation}
where $A$ is the truncation order of the $a$ series. 
The independent expectation values are associated with $n=0,1,2,3$. 
The expansion coefficients $f_{n,0}^{(j)}$ are constrained by the regularity assumption and the differential equations \eqref{pde0}. 
We choose the independent set of free parameters as
\begin{equation}\label{fseta}
    \left(E,\,f_{0,0}^{(0)},\,f_{1,0}^{(0)},\,f_{0,0}^{(1)}\right).
\end{equation}
Since $f_{n,0}^{(0)}=f_{n,0}$, the normalization condition \eqref{normalization} implies that 
\begin{equation}
f_{0,0}^{(0)}=1\,.
\end{equation}
The last two parameters can be expressed as
\begin{equation}
    f_{1,0}^{(0)}=f_{1,0}=\langle{e^{ix}}\rangle=\langle{\cos x}\rangle,\quad f_{0,0}^{(1)}=i f_{0,0,1}=i \langle{p}\rangle\,.
\end{equation}
Accordingly, $f_{1,0}^{(0)}$ should be a real number, while $f_{0,0}^{(1)}$ is purely imaginary. 
In \cite{Tchoumakov:2021mnh,Berenstein:2021loy,Aikawa:2021eai}, the relation between $E$ and $\langle{e^{ix}}\rangle$ was studied using positivity constraints. 
In Sec. \ref{lne0}, we express $\langle{e^{ix}}\rangle$ as a rational function of $E$ using the $1/n$ series and matching conditions. 
Some explicit solutions for the small $a$ expansion coefficients are
\begin{equation}
f_{0,0}^{(2)}=f_{1,0}^{(0)}-\frac E 2 \,,\quad
f_{0,0}^{(3)}=-\frac E 6  f_{0,0}^{(1)}\,,
\quad
f_{1,0}^{(1)}=-\frac{i}{2} f_{1,0}^{(0)}\,,\quad
f_{1,0}^{(2)}=-\frac{2 E+1}{12} f_{1,0}^{(0)}+\frac{1}{3}\,,
\end{equation}
\begin{equation}
  f_{2,0}^{(0)}=\frac{4 E-1}{6} f_{1,0}^{(0)}-\frac{1}{3}\,,\quad
f_{3,0}^{(0)}=\frac{(2E+1)(4E-7)}{15} f_{1,0}^{(0)}-\frac{4 (E-1)}{15}\,.
\end{equation}
For consistency, we verify that 
the solutions to the generic $a$ recursion relation \eqref{recrelation} also become regular in the $a\rightarrow 0$ limit. 
In the small $a$ expansion, $f_{-3,a}$ does not lead to additional independent parameters 
due to $f_{-n,0}^{(0)}=f_{n,0}^{(0)}$ from the $a=0$ recursion relation \eqref{a0rela}. 

Let us emphasize that $\langle{p}\rangle$ is also a free parameter. 
\footnote{The case $n=0$ is nonperturbative in the large $n$ expansion, 
so the independent parameter $f_{0,0,1}=\langle{p}\rangle$ can be absent in the parameter set \eqref{generic-a-large-n} for the large $n$ series. }
For generic $a$, we can express $f_{0,a,1}=\langle{e^{iap}p}\rangle$ in terms of $f_{n,a}$ and $E$, 
as in the derivation of the differential equations \eqref{pde0}. 
However, this expression diverges in the $e^{ia}\rightarrow 1$ limit due to a vanishing denominator.  
According to l'Hôpital's rule, we need to take a derivative of the numerator with respect to $a$
and thus $\langle{p}\rangle$ cannot be expressed in terms of zeroth-order coefficients. 
Alternatively, if we set $a=0$ in the recursion relations \eqref{rela1}, \eqref{rela2} and \eqref{rela3}, 
then $\langle{p}\rangle$ disappears due to vanishing coefficients. 
The fact that $\langle{p}\rangle$ is a free parameter in the $a=0$ recursion relations was also noted in \cite{Aikawa:2021eai}. 
\footnote{In \cite{Tchoumakov:2021mnh}, it was stated that $\langle{p}\rangle=0$, but this is true only in special cases. 
See Fig. \ref{kvs} and Fig. \ref{pvs} for $\langle{p}\rangle$ as functions of $k, E$.}  

Similarly, we can study the divergences in the  $a\rightarrow 2\pi$ limit using the small $(a-2\pi)$ expansion
\begin{equation}\label{smallapi} 
    f_{n,a} \sim \sum_{j=0}^{A} f_{n,2\pi}^{(j)} (a-2\pi)^j,\quad n=0,1,2,3\;,
\end{equation}
where $A$ is the truncation order of the $(a-2\pi)$ series. 
Bloch's theorem indicates 
\begin{equation}
    f_{n, 2\pi}^{(j)} =e^{2\pi i k} f_{n, 0}^{(j)}\,,
\end{equation}
so we can extract the Bloch momentum $k$ by solving the differential equations \eqref{pde0} 
from $a=0$ and $a=2\pi$. 
Let us also mention that the regularity of the $a\rightarrow 2\pi$ limit does not give rise to additional constraints. 
In other words, the divergences at $a=2\pi$ are absent if the $a\rightarrow 0$ limit is regular 
due to a symmetry of $f_{n,a}$. 
\footnote{A precise relation can be found in \eqref{2pi-a}.}

\section{Bootstrap results at special $a$ from matching conditions}\label{secBRa}
In the previous Section, we derived the recursion relations for $f_{n,a}$ in $n$ 
and obtained the $1/n$ series for generic $a$.  
In this Section, we focus on the special cases of $a=0$ and $a=\pi$. 
The first case $a=0$ does not involve the translation operator, 
and was studied numerically using positivity constraints in  \cite{Tchoumakov:2021mnh,Berenstein:2021loy,Aikawa:2021eai}. 
The second case $a=\pi$ is associated with a translation of half period, 
whose recursion relation also exhibits additional simplicity. 
Below, we combine the large $n$ expansion method in Sec. \ref{Lnex} with matching conditions,  
and derive the approximate solutions for $f_{n,0}=\langle{e^{inx}}\rangle$ and $f_{n,\pi}=\langle{e^{inx}e^{i\pi p}}\rangle$ at finite $n$.  

Let us give a brief overview of the matching conditions. 
The recursion relation in $n$ can be solved nonperturbatively at finite $n$ or perturbatively in $1/n$. 
There may exist an overlap region where both approaches are applicable:
\begin{enumerate}[(i)]
\item
If $n$ is not too large, then explicit nonperturbative solutions are doable.
\item
If $n$ is large enough, then perturbative $1/n$ series are accurate.
\end{enumerate} 
As in \cite{Li:2023ewe,Li:2024rod}, we can impose some matching conditions for the two types of solutions:
\begin{equation}
    \langle{e^{inx}e^{ia p}}\rangle^{\text{non-perturbative}}=\langle{e^{inx}e^{ia p}}\rangle^{\text{perturbative}}\,,\quad n\approx M\,,
\end{equation}
where $M$ denotes the matching order. 
Using these additional constraints, we can determine some free parameters in the two types of solutions. 
The resulting nonperturbative solutions at finite $n$ take the forms of rational functions in $E$.  

\subsection{Without translation: $a=0$}\label{lne0}
As a second-order difference equation, 
the $a=0$ recursion relation \eqref{a0rela} admits two types of asymptotic behaviors at large $n$. 
We consider the decaying case
\begin{equation}\label{hoc0}
    f_{n,0}\sim (-4)^{n}\frac{\Gamma\left(n+\frac 1 2\right)}{\sqrt{\pi}\,\Gamma(n+1)^3}\tilde \lambda_0 \left( 1+\sum_{j=1}^{N} \tilde c_0^{(j)} n^{-j} \right)\quad (n\rightarrow \infty),
\end{equation}
and set the prefactor of the growing case $(-4)^{-n}\Gamma(n)^3/\Gamma\left(n+\frac 1 2\right)$ to zero. 
\footnote{For some reason, if we use the growing solution to solve the matching conditions, 
we still obtain an approximate relation between $E$ and $f_{1,0}$, which is less accurate.} 
The explicit coefficients at low orders are 
\begin{equation}
    \tilde c_0^{(1)}=-4 E\,,\quad \tilde c_0^{(2)}=2E(4E+1)\,,\quad \tilde c_0^{(3)}=-\frac{2}{3} \left( 16E^3+16E^2+E+8 \right)\,,
\end{equation}
which are slightly simpler than those in \eqref{hoc}. 
In principle, the large $n$ series can be multiplied by some periodic function that reduces to $1$ at integer $n$. 
The free parameters in the non-perturbative and perturbative solutions are
\begin{equation}
(E, \quad f_{0,0}=\langle{1}\rangle,\quad f_{1,0}=\langle{e^{ix}}\rangle,\quad \tilde \lambda_0)\,.
\end{equation}
The normalization condition $f_{0,0}=1$ in \eqref{normalization} fixes one parameter, 
so there remain three free parameters. 

\begin{figure}
    \includegraphics[scale=0.7]{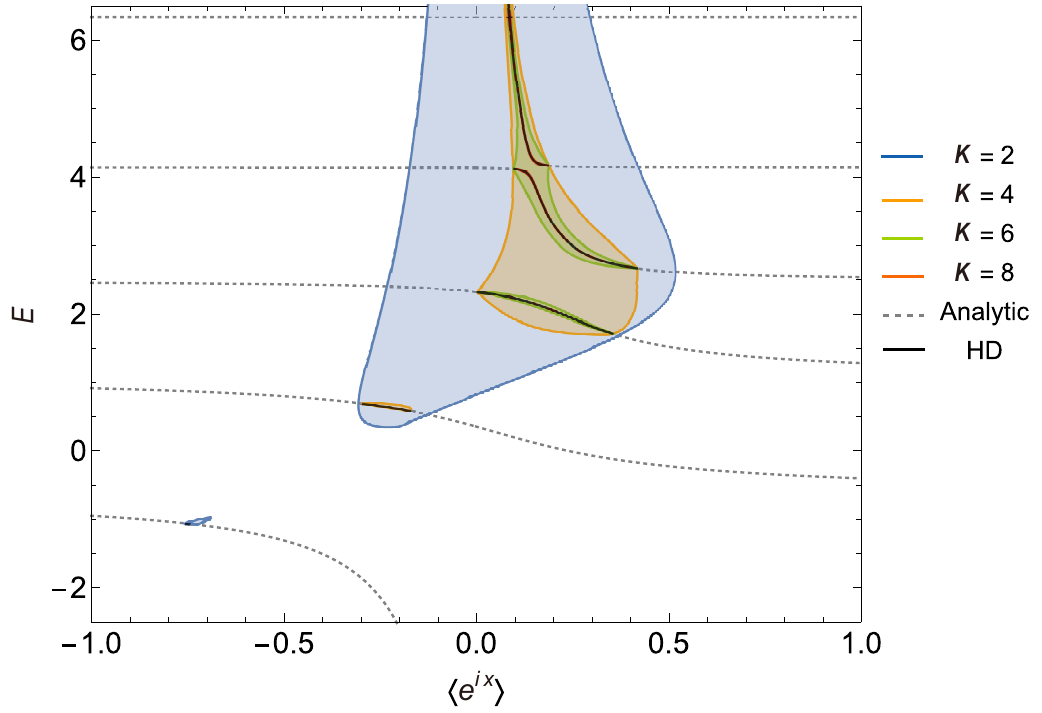}
    \centering
    \caption{The relation between $E$ and $f_{1,0}=\langle{e^{ix}}\rangle$. 
    We use dashed curves to denote our analytic bootstrap solution for $f_{1,0}$,  
    which is a rational function of $E$ from the matching procedure. 
    The $1/n$ series truncation order is $N=2$ and the matching order is $M=6$. 
    We use solid curves to indicate the result from the Hamiltonian diagonalization (HD) method.  
    The colored regions are the bootstrap bounds from positivity constraints, 
    where $K$ denotes the truncation order of the positive semidefinite matrix \cite{Tchoumakov:2021mnh}. 
    In energy bands, our analytic results match well with those of the Hamiltonian diagonalization and bootstrap bounds with $K=8$.  
    The difference is that our analytic solution extends into energy gaps. 
    In Sec. \ref{secBR}, we use some reality conditions to determine the accurate range of energy bands. }
    \label{f100En}
\end{figure}

To solve for $f_{1,0}$ and $\lambda_0$, we consider two matching conditions 
\begin{equation}\label{lneEf}
    f_{M,0}^{\text{(n.p.)}} = f_{M,0}^{\text{(p.)}},\quad f_{M+1,0}^{\text{(n.p.)}} = f_{M+1,0}^{\text{(p.)}}\,,
\end{equation}
where $M$ is the matching order. 
As a concrete example, let us set the truncation order for the $1/n$ series as $N=0$ and the matching order as $M=4$. The solutions of the matching conditions are
\begin{equation} \label{f10-N0-M4}
f_{1,0}=\frac 1 2\frac{E^3-\frac {886}{125}E^2+\frac {12661}{1000}E-\frac {14467}{4000}}
{E^4-\frac {3669}{500}E^3+\frac {3327}{250}E^2+\frac {67}{1000}E-\frac {122597}{16000}}\,,
\end{equation}
\begin{equation}
\tilde\lambda_0=\frac 9 4\frac{1}{E^4-\frac{3669}{500}E^3+\frac {3327}{250}E^2+\frac {67}{1000}E-\frac{122597}{16000}}\,.
\end{equation}
If we substitute the $k=0$ energy in the first band, i.e., $E\approx -1.07013$, into these solutions, 
we obtain $f_{1,0}\approx -0.74419$ and $\lambda_0\approx 0.12633$. 
For comparison, the result from the Hamiltonian diagonalization is $f_{1,0}\approx -0.74415$,
so the relative error is only around $0.005\%$. 
In fact, the simple expression \eqref{f10-N0-M4} provides a good approximation for the relation between $E$ and $\langle{e^{ix}}\rangle$ in the range $E<4$, 
which covers the first three energy bands.

\begin{figure}
    \centering
    %\subfloat[First state with $k=0$ (real part)]
    \begin{subfigure}{.49\linewidth}
        {\includegraphics[width=0.975\linewidth]{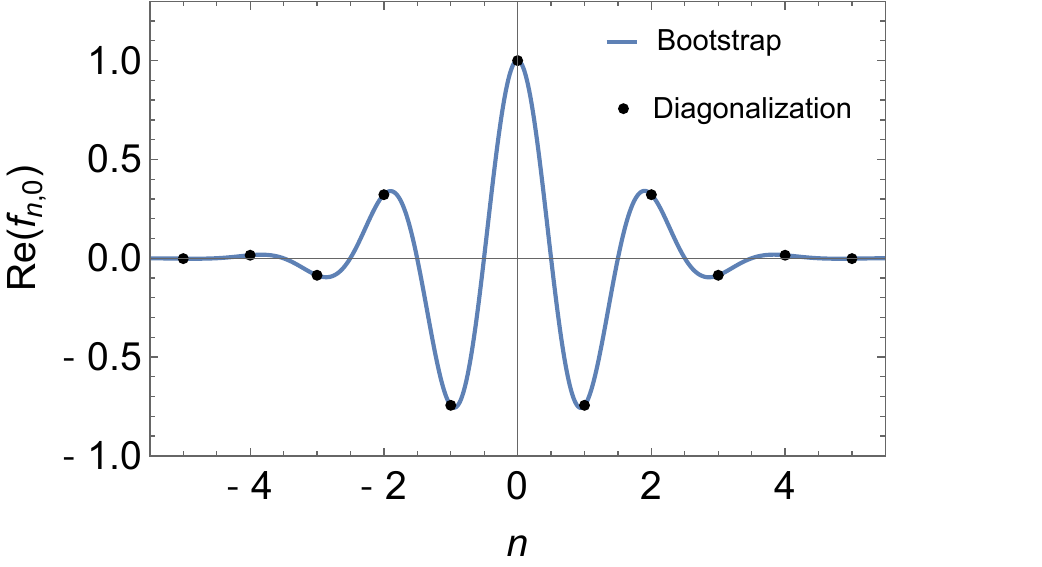}}
        \caption{}
    \end{subfigure}
    %\subfloat[First state with $k=0$ (imaginary part)]
    \begin{subfigure}{.49\linewidth}
        {\includegraphics[width=0.975\linewidth]{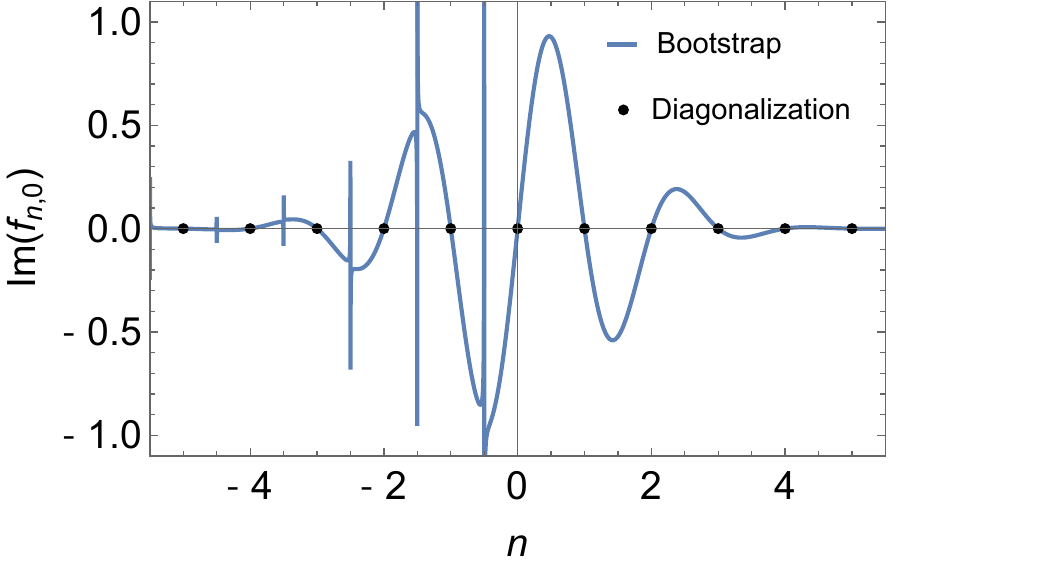}}\\
        \caption{}
    \end{subfigure}
    %\subfloat[Second state with $k=0$ (real part)]
    \begin{subfigure}{.49\linewidth}
        {\includegraphics[width=0.975\linewidth]{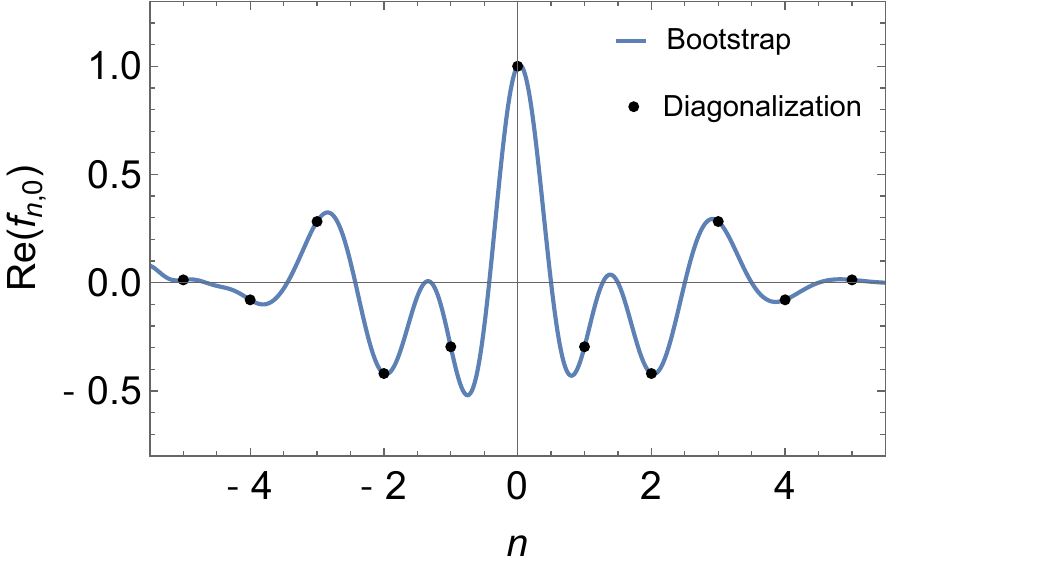}}
        \caption{}
    \end{subfigure}
    %\subfloat[Second state with $k=0$ (imaginary part)]
    \begin{subfigure}{.49\linewidth}
        {\includegraphics[width=0.975\linewidth]{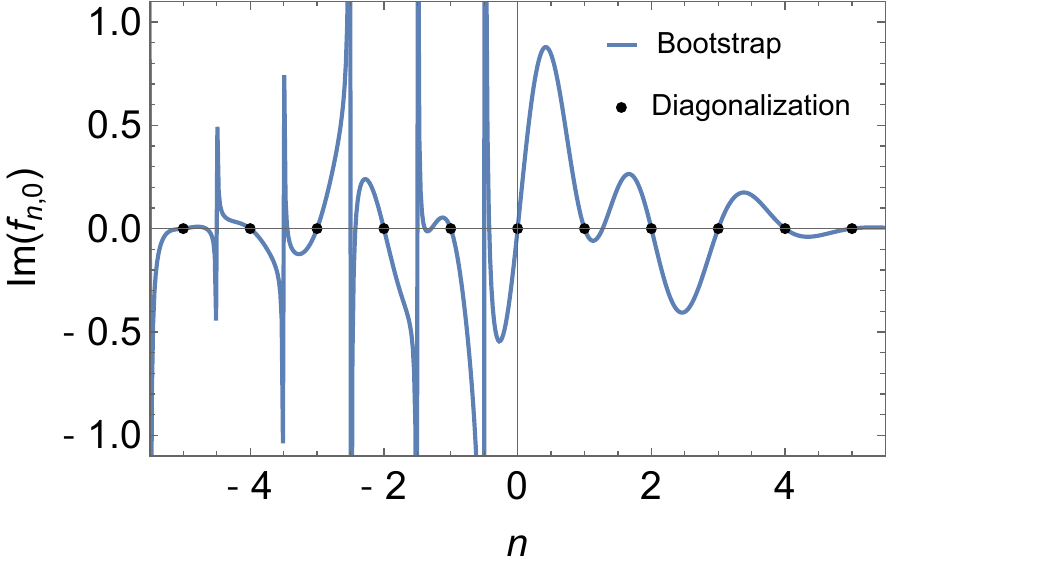}}\\
        \caption{}
    \end{subfigure}
    \caption{Continuous $n$ dependence of $f_{n,0}=\langle{e^{inx}}\rangle$ from the $1/n$ series \eqref{hoc0} and recursion relation \eqref{a0rela} at $k=0$.
    The black dots denote the Hamiltonian diagonalization results. 
    (a) First state with $k=0$ (real part).
    (b) First state with $k=0$ (imaginary part).
    (c) Second state with $k=0$ (real part).
    (d) Second state with $k=0$ (imaginary part).
    }
    \label{fn00}
\end{figure}

In general, the solutions for $f_{1,0}$ and $\tilde \lambda_0$ are given by rational functions of $E$. 
For a fixed truncation order $N$, the solutions improve with the matching order $M$. 
For a large enough $M$, the accuracy of the solutions also increases with $N$. 
In Fig. \ref{f100En}, we compare the result for $N=1, M=6$ 
with those from the Hamiltonian diagonalization 
and the positive bootstrap \cite{Tchoumakov:2021mnh}. 
In the energy bands, our analytic solution matches well with the accurate results from the standard diagonalization method and the positive bootstrap with truncation order $K=8$. 
However, our results for $f_{1,0}$ are not restricted to the physical range, 
as the curves extend into the energy gaps. 
To identify the energy bands, 
we impose reality conditions on some physical quantities, 
which is discussed in Sec. \ref{secBR}. 

We briefly describe the details for obtaining Fig. \ref{f100En}.
In our bootstrap computation, the relation between $E$ and $f_{1,0}$ shown in Fig. \ref{f100En} can be obtained analytically.
In the positive bootstrap computation,
the sizes of the positive semidefinite matrices are $6\times6,\, 10\times10,\, 14\times14,\, 18\times18$ for $(K,L)=(2,1),\, (4,1),\, (6,1),\, (8,1)$ in \eqref{Oeixp}, respectively.
The positive bootstrap analysis is implemented in \textit{Mathematica} using \texttt{PositiveSemidefiniteMatrixQ} and \texttt{RegionPlot},
with the option \texttt{PlotPoints} set to 1000 in \texttt{RegionPlot}.
The plotting range for $\langle e^{ix} \rangle$ is $[-1,1]$ and that for $E$ is $[-1,6]$.
\footnote{For the free parameter $\langle p \rangle$ in the matrices, we consider the case of $\langle p \rangle = 0$ and the cases of $\langle p \rangle \neq 0$.
We find that the bounds for $\langle p \rangle \neq 0$ are inside those of $\langle p \rangle = 0$.}
The computational time increases from approximately 15 minutes to about 40 minutes as $K$ becomes larger.
\footnote{All computations were performed on an Intel i5-13400 processor.}
In the Hamiltonian diagonalization method, 
an eigenfunction of the Hamiltonian is given by 
\begin{equation}\label{diagH}
    \psi  (x)=e^{i k x} \sum_{m=-D}^{D} b_{m}\,  e^{i m x}\,,
\end{equation}
where $D$ is the truncation order, and $b_m$'s are real coefficients. 
The Hamiltonian diagonalization with truncation order $D=8$ takes about 0.05 seconds per point
on the black solid curves in Fig. \ref{f100En}.
\footnote{
To generate the Hamiltonian diagonalization results in Fig. \ref{f100En}, 
we diagonalize the Hamiltonian for a given $k$ to obtain the eigenvalues $E$ and the corresponding eigenfunctions,
and then evaluate the expectation value $f_{1,0}$ for each eigenstate.
}

Although $f_{n,0}$ is originally defined at integer $n$, 
we can extend $n$ to noninteger values using \eqref{hoc0}, which are reliable for large positive $n$. 
For small positive $n$ and negative $n$, we use the recursion relation \eqref{a0rela} 
to write $f_{n,0}$ in terms of those with greater $n$ so that \eqref{hoc0} is applicable. 
In Fig. \ref{fn00}, we compare our results for $f_{n,0}$ with those from the Hamiltonian diagonalization, 
which match well for integer $n$ even in the negative range. 
For $n<0$, the general $n$ results for $f_{n,0}$ exhibit singular behavior at $n=-\frac{1}{2},\,-\frac{3}{2},\,\dots$.
According to the recursion relation \eqref{a0rela}, $f_{n-1,0}$ stays finite in the $n\rightarrow \frac 1 2 $ limit only if the sum of the other two terms is zero.
However, their sum is finite and purely imaginary, 
which leads to the divergences in the imaginary part of $f_{n,0}$ at $n=-1/2$ and other negative half-integer values.

\subsection{Half-period translation: $a=\pi$}\label{lnepi}
We can also use the large $n$ expansion and matching conditions to solve for $f_{n,\pi}$. 
According to \eqref{hoc}, \eqref{cpmja}, \eqref{xi2pi} and \eqref{bkb}, 
the large $n$ expansion of $f_{n,\pi}$ reads
\begin{equation}\label{hocpi}
    f_{n,\pi} \sim e^{2 n} n^{-2 n} n^{-\frac 3 2} \left((2i)^n  + (-2i)^n  e^{2\pi i k } \right) 
    \lambda_1\left( 1+ \sum_{j=1}^{N}c_{+,\pi}^{(j)}\,n^{-j} \right),
\end{equation}
where $N$ is the truncation order of the $1/n$ series.
We can also use $f_{n,\pi}$ as the boundary conditions of the differential equations and determine the properties of the Bloch bands.

\begin{figure}
    \includegraphics[width=0.75\columnwidth]{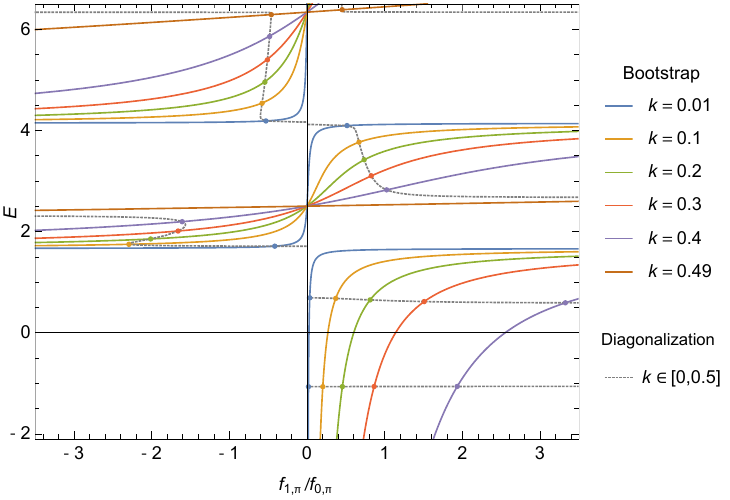}
    \centering 
    \caption{Relation between $E$ and  $f_{1,\pi}/f_{0,\pi}=\langle e^{ix}e^{i\pi p} \rangle / \langle e^{i\pi p} \rangle$ at various $k$. 
    The solid curves are our bootstrap results, 
    where the $1/n$ series truncation order is $N=2$ and the matching order is $M=17$. 
    The dashed curves represent the Hamiltonian diagonalization results.
    As expected, the two types of curves intersect at the solid points when they have the same $k$.  
    }
    \label{f1pi0Enk}
\end{figure}

As in Sec. \ref{lne0}, we derive the relation between $E$ and $f_{n,\pi}$ from the matching conditions.
For $n\geq 0$, the free parameters are
\begin{equation}
    (E, \,f_{0,a},\,f_{1,a},\,f_{2,a},\, f_{3,a},\,k,  \lambda_1 ).
\end{equation}
In principle, $\lambda_1$ is fixed by the normalization condition \eqref{normalization} 
after solving the differential equations \eqref{pde0}. 
To determine $(f_{0,a},\,f_{1,a},\,f_{2,a},\, f_{3,a})$, we solve the matching conditions 
\begin{equation}\label{lneEfpi}
    f_{n,\pi}^{\text{(n.p.)}} = f_{n,\pi}^{\text{(p.)}},\quad n=M, M+1, M+2, M+3\,.
\end{equation}
In Fig. \ref{f1pi0Enk}, we present the relation between $E$ and $f_{1,\pi}/f_{0,\pi}$ for $k=0.01, 0.1, 0.2, 0.3, 0.4, 0.49$, where the truncation parameters are $N=2, M=17$. 
\footnote{As the $f_{n,\pi}$ vanishes for even $n$ at $k=1/2$, the combination $f_{1,\pi}/f_{0,\pi}$ diverges in the $k\rightarrow 1/2 $ limit. Therefore, we only present the result for $k=0.49$. On the other hand, $f_{n,\pi}$ vanishes for odd $n$ at $k=0$, so we only consider the case of $k=0.01$.  The large matching order $M$ is to ensure the accuracy of $E$ in the fifth band. } 
The ratio $f_{1,\pi}/f_{0,\pi}$ does not depend on the prefactor $\lambda_1$. 
According to the identity \eqref{ap-identity}, $e^{-i\pi k}f_{n,\pi}$ is a real number, 
so the ratio $f_{1,\pi}/f_{0,\pi}$ should be real. 
We also present the Hamiltonian diagonalization results as functions of $k$, 
which intersect with the constant-$k$ curves at the correct energies. 

\begin{figure}
    \centering
    %\subfloat[First state with $k=0$]
    \begin{subfigure}{.49\linewidth}
        {\includegraphics[width=0.975\linewidth]{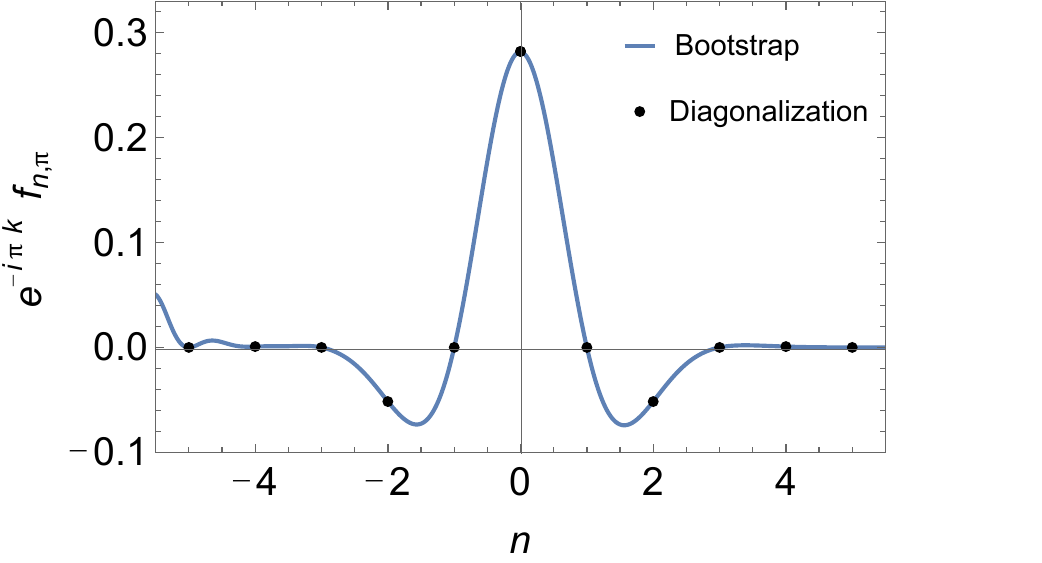}}
        \caption{}
    \end{subfigure}
    %\subfloat[Second state with $k=0$]
    \begin{subfigure}{.49\linewidth}
        {\includegraphics[width=0.975\linewidth]{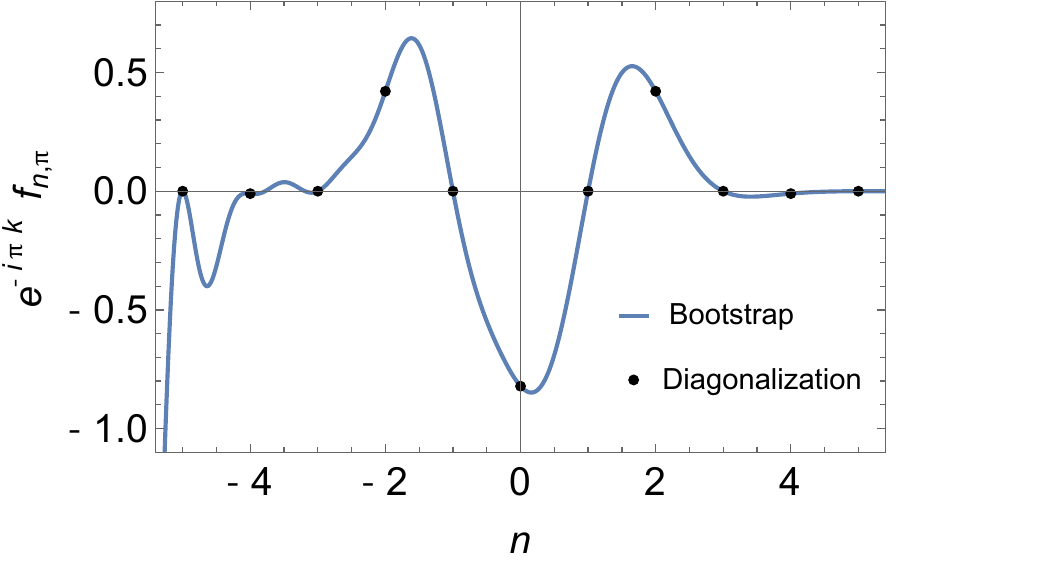}}\\
        \caption{}
    \end{subfigure}
    %\subfloat[First state with $k=0.1$]
    \begin{subfigure}{.49\linewidth}
        {\includegraphics[width=0.975\linewidth]{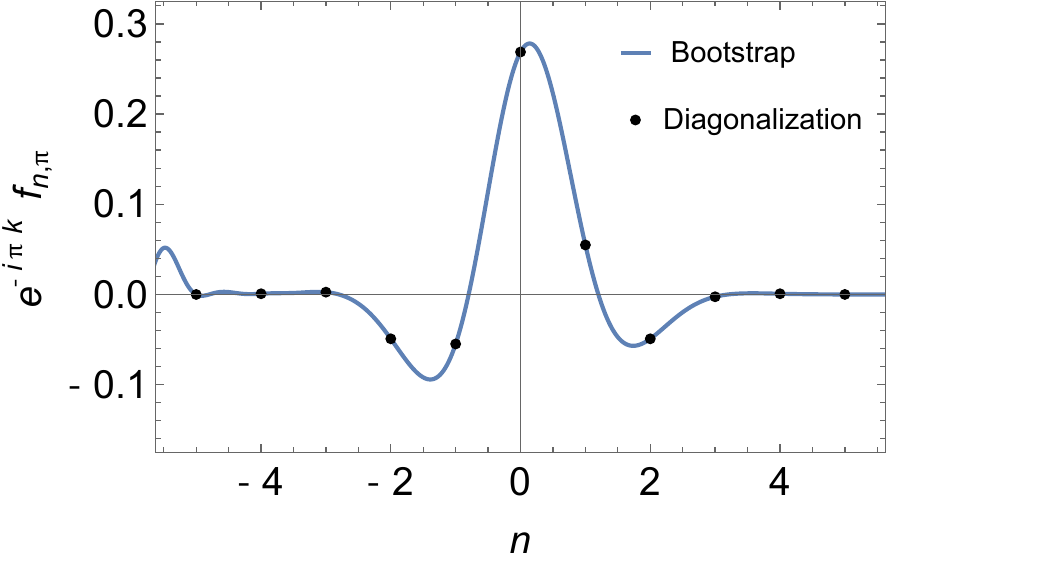}}
        \caption{}
    \end{subfigure}
    %\subfloat[Second state with $k=0.1$]
    \begin{subfigure}{.49\linewidth}
        {\includegraphics[width=0.975\linewidth]{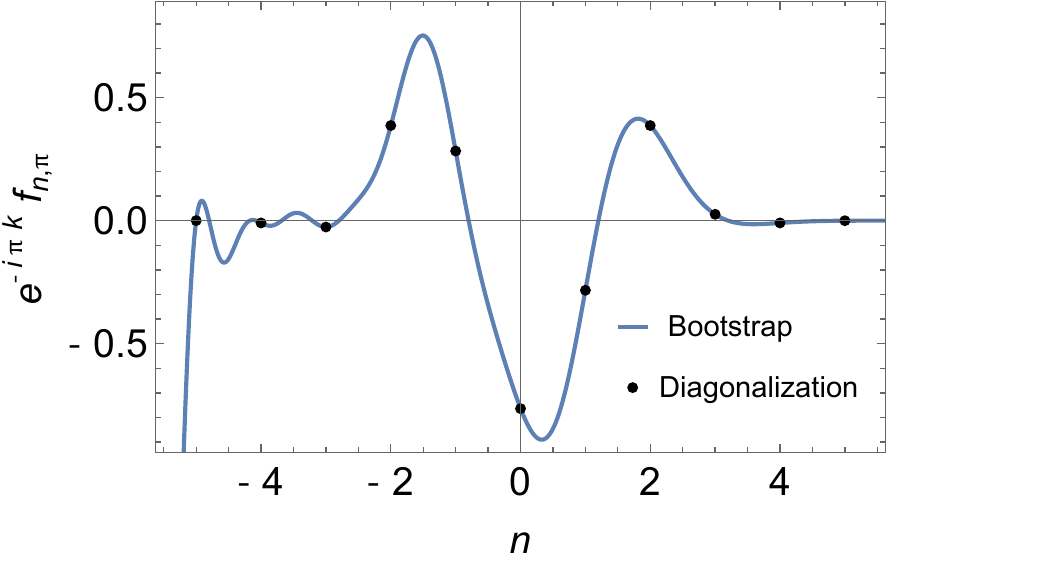}}\\
        \caption{}
    \end{subfigure}
    \caption{Continuous $n$ dependence of $e^{-ik\pi}f_{n,\pi}=\langle e^{inx}e^{i\pi(p-k)} \rangle$ for $k=0, 0.1$. 
    The blue curves represent the results from the matching procedure 
    with the $1/n$ series truncation order $N=5$ and the matching order $M=11$. 
    The black dots are computed from the Hamiltonian diagonalization.
    (a) First state with $k=0$.
    (b) Second state with $k=0$.
    (c) First state with $k=0.1$.
    (d) Second state with $k=0.1$.
    }
    \label{fnpi0}
\end{figure}

We can also study the continuous $n$ dependence of $e^{-i\pi k}f_{n,\pi}$ 
using the $1/n$ series \eqref{hocpi} and the recursion relation \eqref{recpi}. 
In Fig. \ref{fnpi0}, we present the bootstrap results for $k=0$ and $k=0.1$.
If $f_{-3,\pi}=-f_{3,\pi}$, then the recursion relation \eqref{recpi} 
implies that 
$f_{-n,\pi}=(-1)^n f_{n,\pi}$ for integer $n$, which is confirmed by the Hamiltonian diagonalization results.

\section{Bootstrap results at generic $a$ from differential equations}\label{secBR}
In this Section, we use the solutions for $f_{n,a}$ with $a=0,\pi$ 
as the boundary conditions. 
After solving the differential equations \eqref{pde0} and 
determining the $a$ dependence of $f_{n,a}$ for $a\in [0, 2\pi]$, 
we obtain the continuous relation between $\langle{p}\rangle$ and $k$ for a given energy $E$. 
The reality of both $\langle{p}\rangle$ and $k$ then plays the role of a quantization condition.  
At most two values in the first Brillouin zone $|k|\leq \frac 1 2$ satisfy these reality requirements. 
In this way, we deduce the relation between $E$ and $k$, i.e., the dispersion relation. 

In Sec. \ref{M1}, we use the $a=0$ solutions from Sec. \ref{lne0} to generate the boundary conditions for the differential equations. 
Then we solve the differential equations numerically 
and use the $\langle{p}\rangle$ dependence of $k$ to extract the real solutions of $k$. 
In Sec. \ref{M2}, we use the $a=\pi$ solutions from Sec. \ref{lnepi} to set the boundary conditions 
and solve the differential equations by truncated Fourier series. 
Then we examine the $k$ dependence of $\langle{p}\rangle$ and use the reality of $\langle{p}\rangle$ to fix $k$. 
In both cases, the divergence issues in the $e^{ia}\rightarrow 1$ limit are resolved by the small $a$ expansion in Sec. \ref{sae}. 

\subsection{Direct numerical solution}\label{M1}
As discussed in Sec. \ref{lne0}, we can readily solve the recursion relation for $f_{n,a}$ at $a=0$ using the large $n$ expansion and matching conditions. 
However, we cannot directly set the boundary at $a=0$ 
because some coefficients of the differential equations \eqref{pde0} would diverge. 
In Sec. \ref{sae}, we introduce the small $a$ expansion approach 
to implement the regularity of the $a\rightarrow 0$ limit. 
Accordingly, the coefficients of the small $a$ expansion are expressed in terms of 
\begin{equation}
\left(E\,,\quad f_{1,0}=\langle{e^{ix}}\rangle \,,\quad f_{0,0,1}=\langle{p}\rangle\right)\,,
\end{equation} 
where $\langle{e^{ix}}\rangle$ can be approximated by a rational function of $E$.  
Let us emphasize that the real parameter $\langle{p}\rangle$ is not constrained by the recursion relation at $a=0$. 

In practice, we use a small value of $a$ as a regulator. 
For instance, we set the left boundary at $a=1/10$ to avoid the explicit divergences 
associated with the $a\rightarrow 0$ limit, 
and evaluate the boundary conditions $f_{n,1/10}$ accurately 
using the truncated $a$ series in \eqref{smallao}. 
At a given energy $E$, we can solve the differential equations numerically 
and obtain a family of solutions parametrized by $\langle{p}\rangle$. 
Similarly, we set the right boundary at $2\pi-1/10$, 
and extract the $a=2\pi$ results, i.e., $f_{n,2\pi}$, using the truncated $(a-2\pi)$ series in \eqref{smallapi}. 
Then the Bloch momentum computed from
\begin{equation}
    k=\frac{\ln f_{0,2\pi}}{2\pi i}=\frac{\ln \langle e^{2\pi i p} \rangle}{2\pi i}
\end{equation}
is a function of $\langle{p}\rangle$. 
We confine our discussion to the case of real $\langle{p}\rangle$.

\begin{figure}
    \includegraphics[width=0.7\columnwidth]{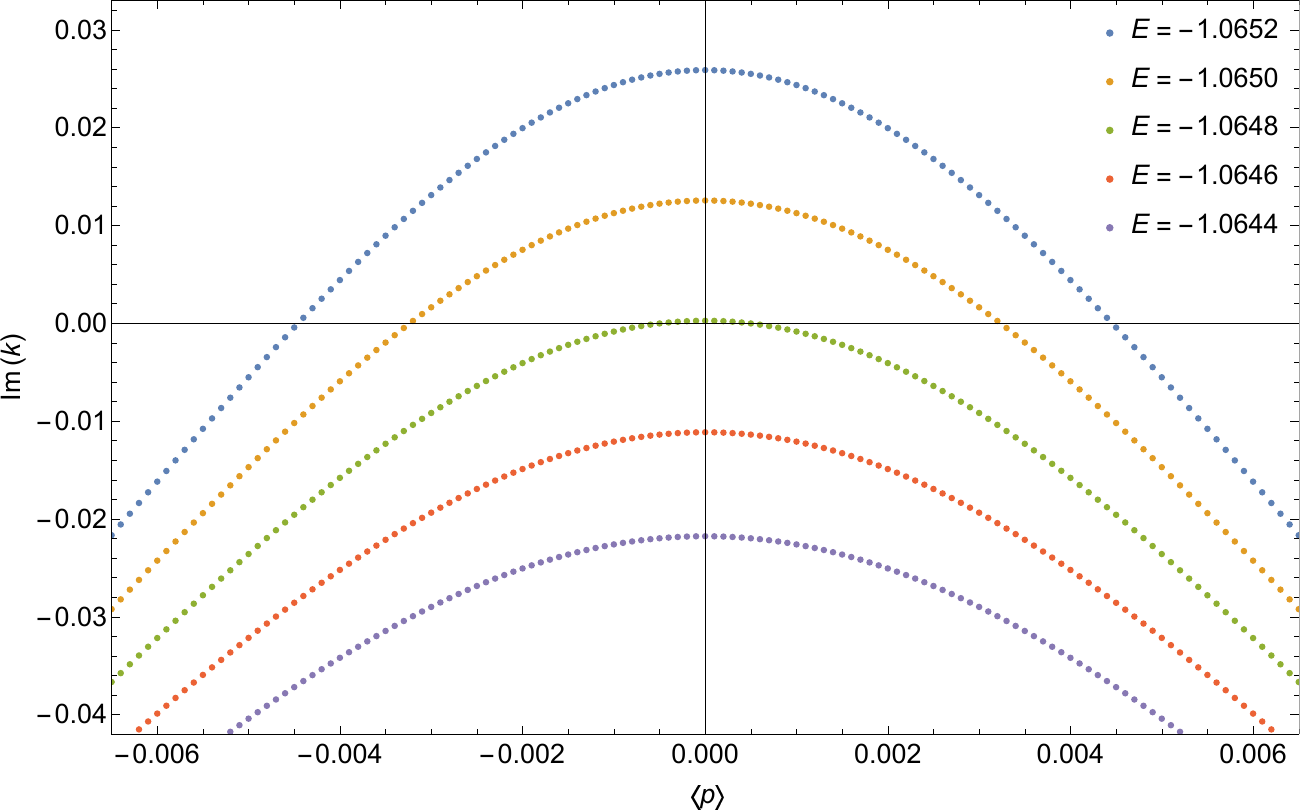}
    \centering 
    \caption{Imaginary part of $k$ as a function of real $\langle{p}\rangle$. 
    We consider the energies around the maximum of the first energy band, i.e., $E_\text{max}^{(1)}\approx-1.064796$. 
    }
    \label{scanp}
\end{figure}

For generic $(E, \langle{p}\rangle)$, the resulting $k$ is a complex number 
with a finite imaginary part. 
However, a physical Bloch momentum should be real. 
For a given $E$, we can use the reality condition 
\begin{equation}\label{reality-k}
\text{Im}(k)=0
\end{equation}
to determine the physical values of $\langle{p}\rangle$ and $k$. 
For illustration, let us consider the cases around $E_{\max}^{(1)}$, i.e., the maximum energy of the first band. 
In Fig. \ref{scanp}, we can see that the imaginary part of $k$ is a function of $\langle{p}\rangle$.   
The energies are chosen to be around  $E_{\max}^{(1)}$. 
If the energy $E$ is slightly below $E_\text{max}^{(1)}$, 
the imaginary part of $k$ has two symmetric zeros at $\langle{p}\rangle\neq 0$, 
and we obtain two physical Bloch momenta. 
As the energy $E$ increases, the two zeros move toward $\langle{p}\rangle=0$ and collide at the maximum energy $E=E_\text{max}^{(1)}$.
If the energy $E$ is slightly above $E_\text{max}^{(1)}$, 
then the reality condition \eqref{reality-k} has no solution at real $\langle{p}\rangle$.

\begin{table}
    \centering
    \begin{tabular}{c|c|c}
    $m$& $E^{(m)}_\text{min}$ & $E^{(m)}_\text{max}$\\
    \hline
    $1$ & -1.070129704575631 & -1.064795725140236 \\
    $2$ & 0.579502042526632 & 0.686720256798165 \\
    $3$ & 1.707268708641598 & 2.315361533026600 \\
    $4$ & 2.667756775880152 & 4.113008822532208 \\
    $5$ & 4.162454726704300 & 6.332636217945250 \\
    \hline
    \end{tabular}
    \caption{Minima and maxima of the energy bands. 
    We use $m$ to label the energy bands. 
    These reference values are determined by the Hamiltonian diagonalization with truncation order $D=8$.
    At this truncation order, the Hamiltonian diagonalization is accurate to more than 20 significant digits 
    when we compute the energies in the first five energy bands. 
    The values in the table are displayed to 15 decimal places.
    To obtain the results in the table, we perform the calculations for $k=0$ and $k=1/2$ in \eqref{diagH}, which correspond to the minimum or maximum of an energy band.
    }
    \label{diagE}
\end{table}
\begin{table}
    \footnotesize
    \centering
    \begin{tabular}{c|ccc|ccc}
    m & \multicolumn{3}{c|}{$\Delta E^{(m)}_\text{min}$}  & \multicolumn{3}{c}{$\Delta E^{(m)}_\text{max}$}\\
    \cline{2-7}
    & $A=7$ & $A=10$ & $A=13$ & $A=7$ & $A=10$ & $A=13$ \\
    \hline
    $1$ & $-7.0 \times 10^{-5}$ & $1.3 \times 10^{-10}$ & $1.5 \times 10^{-14}$ & $-7.2\times 10^{-5}$ & $1.3 \times 10^{-10}$ & $2.3 \times 10^{-14}$ \\
    $2$ & $2.6 \times 10^{-5}$ & $-3.6 \times 10^{-11}$ & $-1.3 \times 10^{-14}$ & $6.0 \times 10^{-5}$ & $-9.5 \times 10^{-11}$ & $-2.4 \times 10^{-14}$ \\
    $3$ & $-1.1 \times 10^{-4}$ & $3.7 \times 10^{-10}$ & $1.2 \times 10^{-13}$ & $3.1 \times 10^{-4}$ & $-1.3 \times 10^{-10}$ & $-2.0 \times 10^{-13}$ \\
    $4$ & $-6.0 \times 10^{-4}$ & $1.5 \times 10^{-9}$ & $7.5 \times 10^{-13}$ & $2.1 \times 10^{-4}$ & $1.1 \times 10^{-9}$ & $1.9 \times 10^{-13}$ \\
    $5$ & $-1.2 \times 10^{-3}$ & $3.2 \times 10^{-9}$ & $1.3 \times 10^{-12}$ & $-6.8 \times 10^{-4}$ & $4.1 \times 10^{-9}$ & $2.4 \times 10^{-12}$ \\
    \hline
    \end{tabular}
    \caption{The differences between our bootstrap results and  {the reference values listed in Table \ref{diagE}.
    The reference values are obtained from HD.}
    The energy difference is defined as $\Delta E^{(m)} \equiv E^{(m)}_{\text{Bootstrap}}-E^{(m)}_{\text{HD}}$.  
    The subscripts indicate the minimum or maximum of the $m$th energy band. 
    We use $A$ to denote the truncation orders of the $a$ series in \eqref{smallao} and $(a-2\pi)$ series in \eqref{smallapi}. }
    \label{allowedbands}
\end{table}

The absence of a solution to the reality condition \eqref{reality-k} with real $\langle{p}\rangle$ 
provides a clear signature for forbidden bands. 
In Table \ref{diagE},  we summarize the reference values for the maximum and minimum energies of the first five energy bands from the standard Hamiltonian diagonalization method. 
In Table \ref{allowedbands}, we present the differences between our bootstrap predictions and the diagonalization results. 
We choose $N=3$ as the $1/n$ series truncation order  and $M=30$ as the matching order.  
In this way, the error from the rational approximation for $\langle{e^{ix}}\rangle$ is negligible. 
The main source of error is associated with the truncated series in $a$ and $a-2\pi$, 
whose truncation orders are the same and denoted by $A$. 
In Table \ref{allowedbands}, we show that 
the accuracy of our bootstrap results improves rapidly with the truncation order $A$.

 {
We briefly describe our computational procedure.
In our bootstrap computation, we employ \texttt{ParametricNDSolve}
\footnote{ {The parameter in \texttt{ParametricNDSolve} is $E$. 
Another free parameter, $\langle p \rangle$, is set to $0$ when computing the minima and maxima of the energy bands.}} 
and \texttt{FindRoot} in \textit{Mathematica}.
We use \texttt{ParametricNDSolve} to solve the differential equations numerically and use \texttt{FindRoot} to determine the values of $E$ for which $\text{Im}(k)=0$.
In this process, each call to \texttt{FindRoot} invokes a \texttt{ParametricFunction} returned by \texttt{ParametricNDSolve}.
Each invocation of \texttt{ParametricFunction} solves the differential equations numerically.
\footnote{ {The computational time of \texttt{ParametricFunction} is approximately independent of the values of $A$ used in the computation.}}
The computational time of a \texttt{ParametricFunction} returned by \texttt{ParametricNDSolve} with a given \texttt{WorkingPrecision} is nearly constant.
The total computational cost also depends on \texttt{FindRoot}, which requires more iterations when a higher \texttt{WorkingPrecision} is used for larger $A$.
}

 {Let us compare the computational cost of the Hamiltonian diagonalization method with that of our bootstrap method.
For the first band, agreement with the reference values in Table \ref{diagE} to 4, 9, and 13 decimal places 
is obtained with truncation orders $D=3,6,7$, in the Hamiltonian diagonalization method.
The corresponding computational times are approximately 0.0007s, 0.0023s, and 0.0033s.
To achieve the same accuracy with our bootstrap method,
we set the truncation orders of the small $a$ expansion and the small $(a-2\pi)$ expansion to $A = 7, 10, 13$, as shown in Table \ref{allowedbands},
and set \texttt{WorkingPrecision} to 25, 35, and 40 in \texttt{ParametricNDSolve}.
With these settings, each call to \texttt{ParametricFunction} takes approximately 0.1s, 0.5s, and 1.0s.
\footnote{ {To achieve the same accuracy using the positive bootstrap method, we set $(K,L) = (5,1),\,(7,1),\,(9,1)$ in \eqref{Oeixp}, respectively.
We use \texttt{SemidefiniteOptimization} in \textit{Mathematica} to solve the SDP problem and take \texttt{MOSEK} for the \texttt{Method} option in \texttt{SemidefiniteOptimization} \cite{Aikawa:2025dvt}.
Each call to \texttt{SemidefiniteOptimization} takes approximately 0.005s, 0.019s, and 0.034s, respectively.}}
The total computational time for finding the minimum or maximum of the band is about 1s, 5s, and 10s.
For the second band, agreement with the reference values to 4, 9, and 13 decimal places
is obtained with truncation orders $D=4,6,7$, in the Hamiltonian diagonalization method.
In the bootstrap computation, we use the same settings as for the first band and find a comparable computational cost.
Since the bootstrap computation is formulated in terms of observables, it requires more computational resources than the Hamiltonian diagonalization method, 
which is based on the wave functions.}

 {
Once the parameters in the differential equations are determined, our approach readily yields the continuous $a$ dependence of the observables.
For comparison, one needs to evaluate a numerical integral at each $a$ in the diagonalization approach. 
In Fig. \ref{avs}, we present the $a$ dependence of $e^{-ika}f_{n,a}$ for the case of $k=0.1$ in the first band.
The factor $e^{-ika}$ is included to simplify the presentation of our results.
Our bootstrap results are well consistent with the Hamiltonian diagonalization results. 
In Sec. \ref{M2}, we will also explain a different method to solve the differential equations, which is based on the truncated Fourier series.
}

\begin{figure}
    \includegraphics[width=0.90\columnwidth]{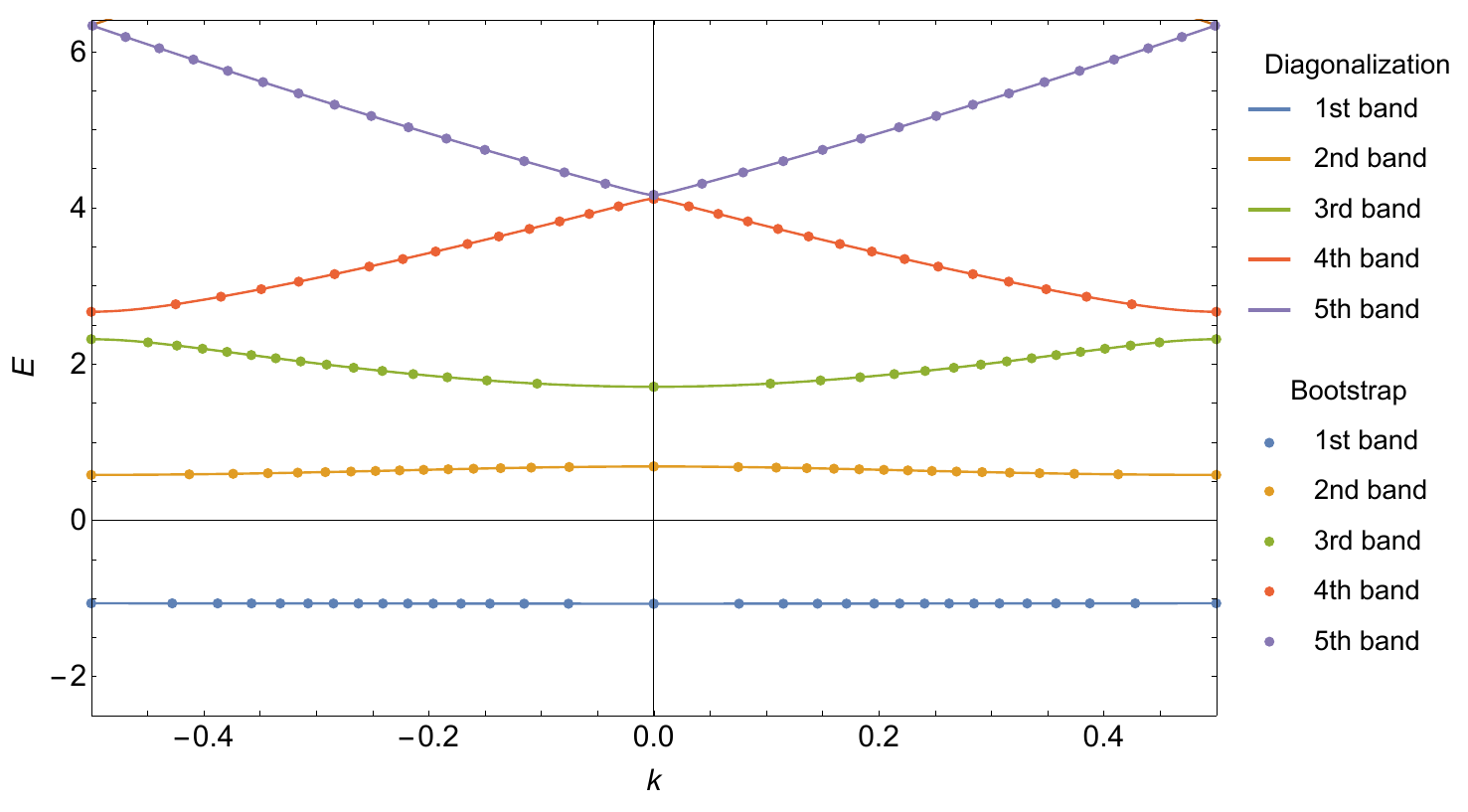}
    \centering
    \caption{Relation between $E$ and $k$ from the direct numerical solutions to the differential equations \eqref{pde0} and reality condition $\text{Im}(k)=0$. 
     We use different colors to indicate different energy bands. 
     The bootstrap results (dots) are in excellent agreement with the Hamiltonian diagonalization results (curves). 
  }
    \label{BB}
\end{figure}

\begin{figure}
    \centering
    %\subfloat[The real part]
    \begin{subfigure}{.99\linewidth}
        {\includegraphics[width=0.9\linewidth]{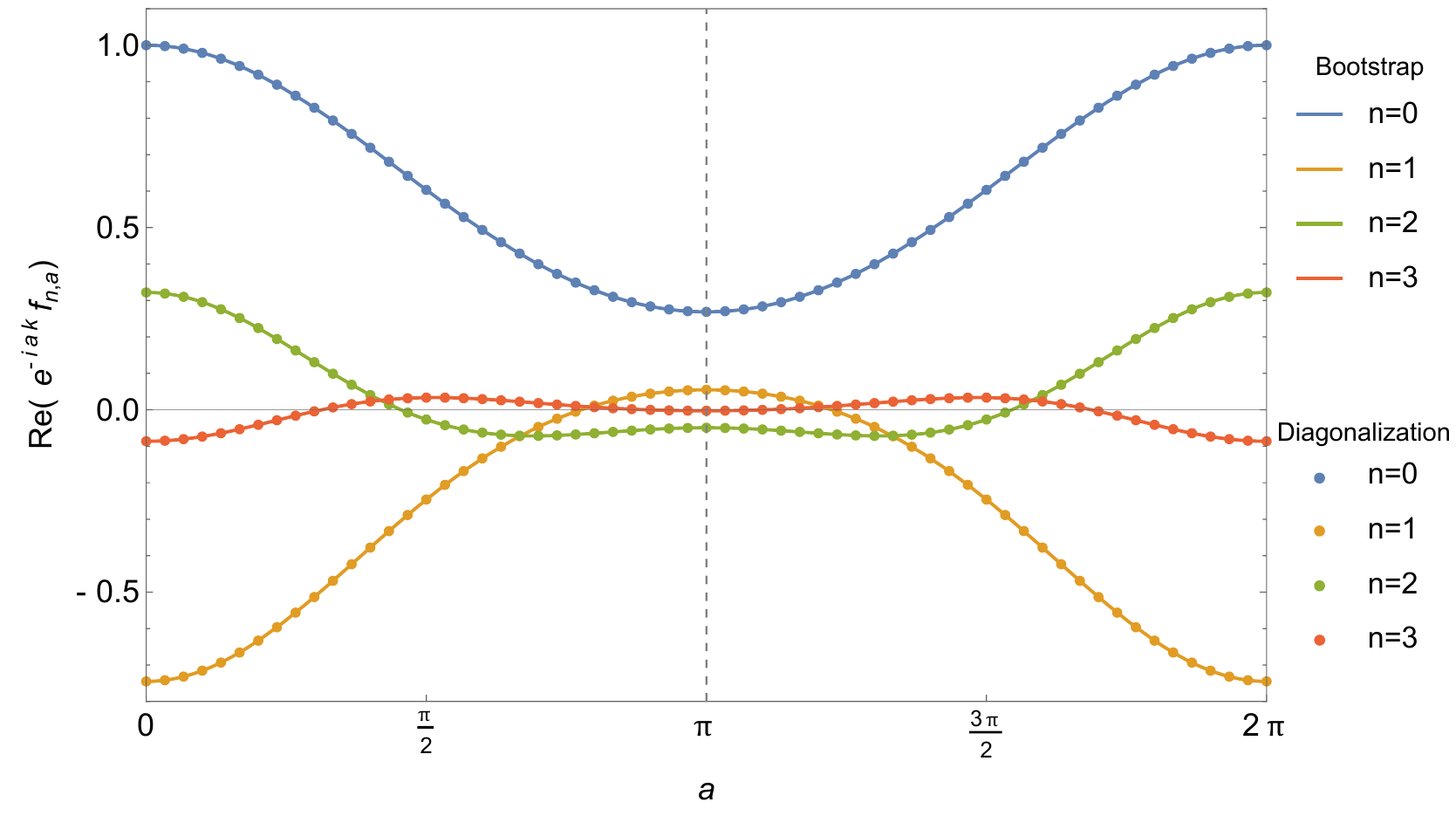}}\quad
        \caption{}
    \end{subfigure}
    %\subfloat[The imaginary part]
    \begin{subfigure}{.99\linewidth}
        {\includegraphics[width=0.88\linewidth]{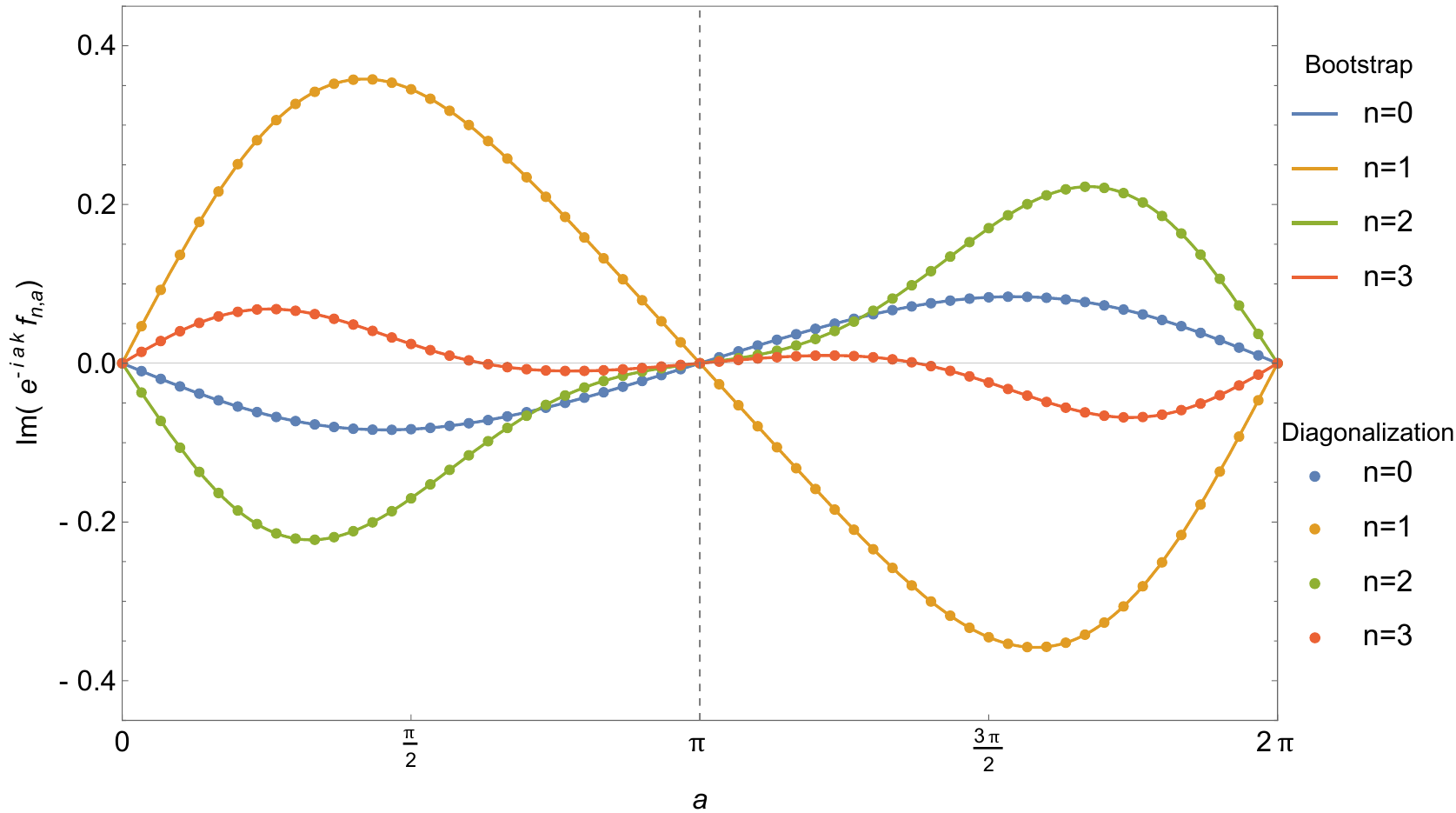}}
        \caption{}
    \end{subfigure}
    \caption{Continuous $a$ dependence of $e^{-ika}f_{n,a}=\langle e^{inx}e^{i(p-k)a} \rangle$ with $n=0,1,2,3$ (blue, orange, green, red).  
    We consider the $k=0.1$ case in the first band with $E\approx-1.0696$. 
    The curves represent our bootstrap results, while the dots are the Hamiltonian diagonalization results. 
    (a) The real part.
    (b) The imaginary part. }
    \label{avs}
\end{figure}

\clearpage

In Fig. \ref{BB}, we present the bootstrap results
for the dispersion relations of the first five energy bands. 
\footnote{ {For the cases of $k \neq 0$ and $|k| \neq 1/2$, 
we choose a value of $E$ in the energy bands and treat $\langle p \rangle$ 
as a free parameter in \texttt{ParametricNDSolve} with \texttt{WorkingPrecision}=30.
We then use \texttt{FindRoot} with \texttt{WorkingPrecision}=10 to determine the values of $\langle p \rangle$ 
satisfying $\text{Im}(k)=0$. 
Each point takes about 10 seconds.}}
The truncation order for the $a$ or $(a-2\pi)$ series is $A=13$. 
We choose an equal spacing for the energies within the allowable bands, 
so the horizontal spacing varies with $k$. 
For $0\leq k \leq \frac 1 2$, the energies of the first, third, and fifth bands are monotonically increasing, 
while those of the second and fourth bands are decreasing. 
Our bootstrap determinations of the Bloch momenta are in excellent agreement with 
the results of the Hamiltonian diagonalization method. 

\begin{figure}
    \centering
    {\includegraphics[width=0.9\linewidth]{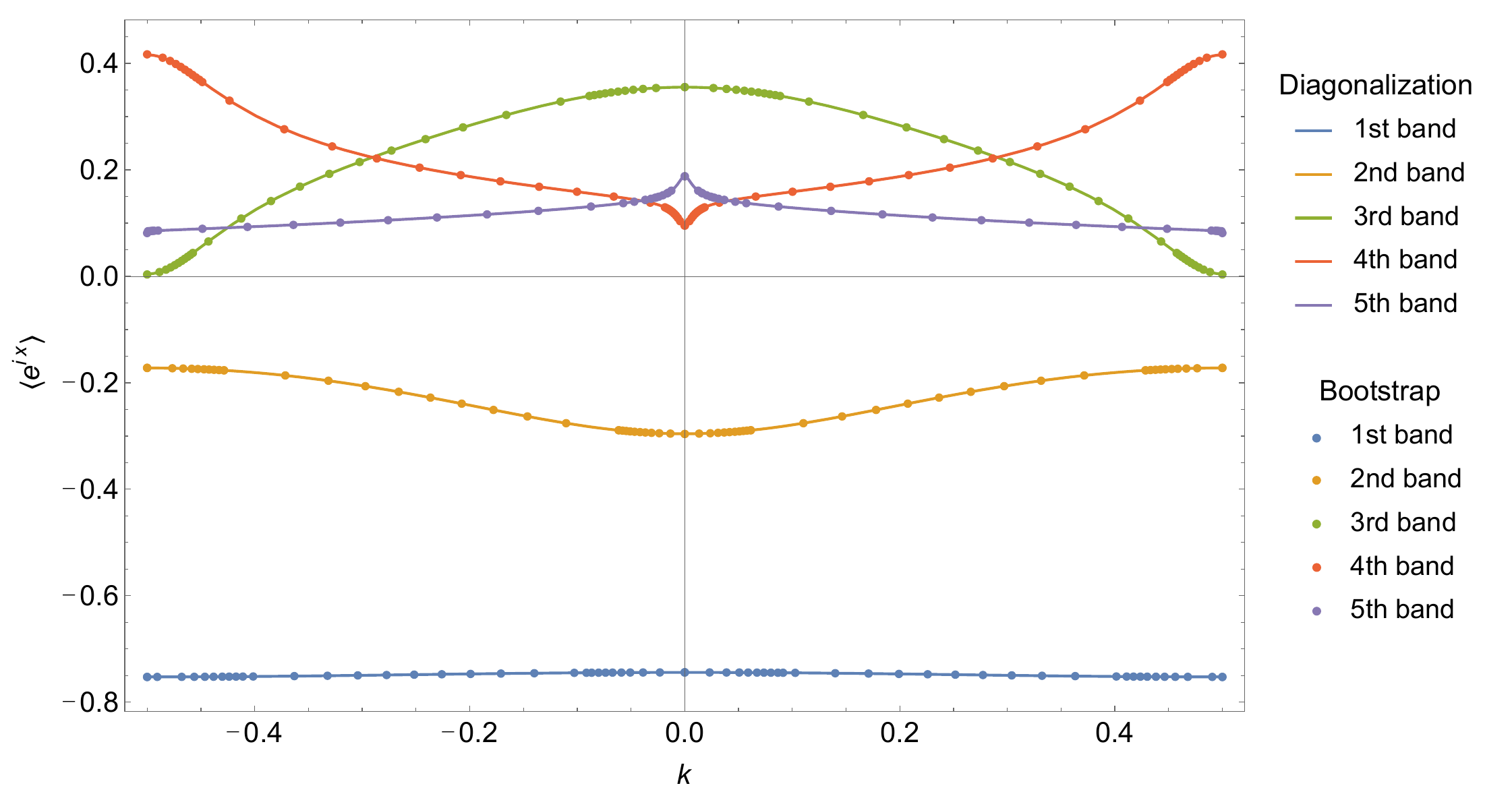}}\quad
    {\includegraphics[width=0.87\linewidth]{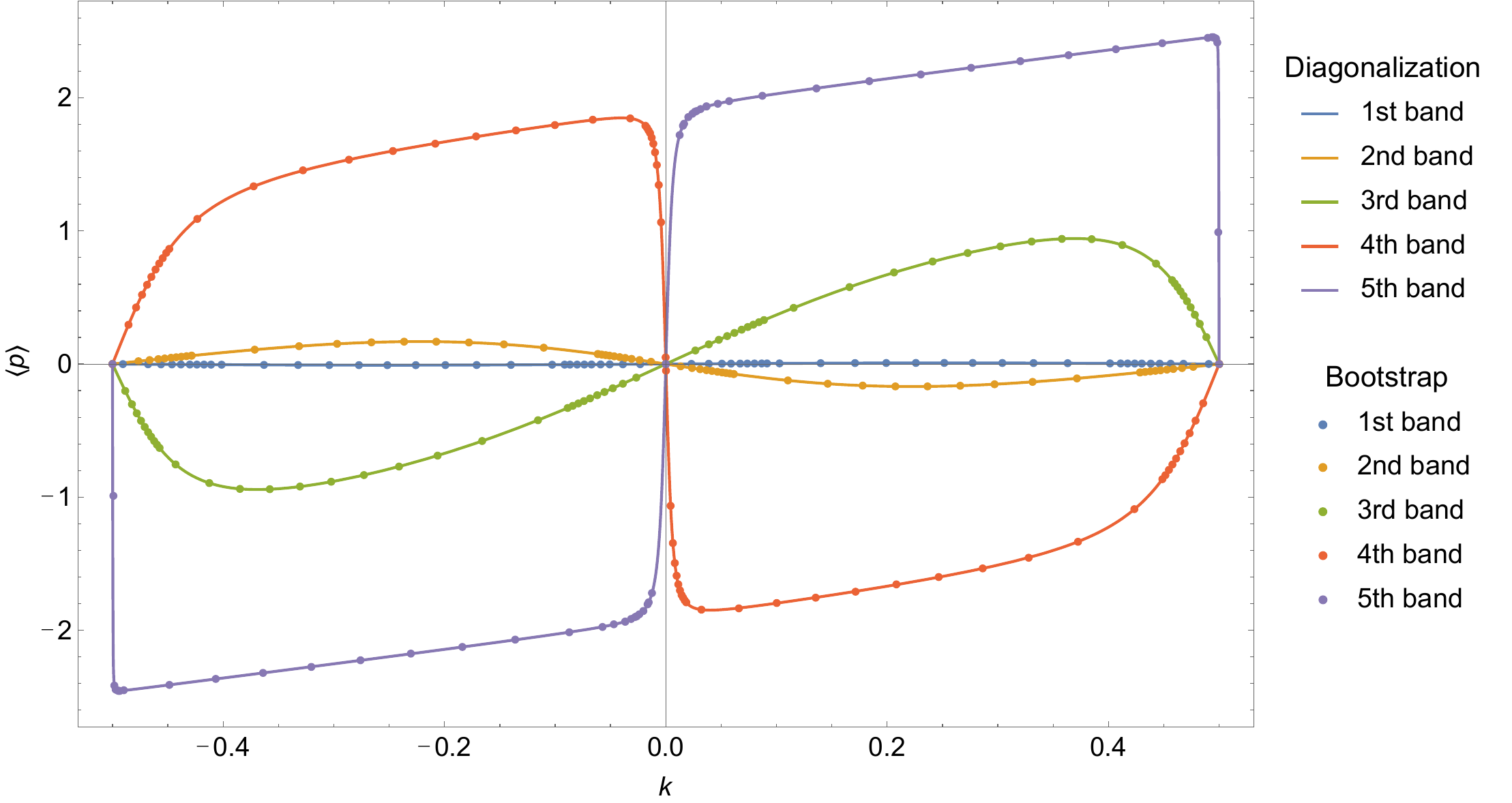}}
    \caption{$k$ dependence of $\langle{e^{ix}}\rangle$ and $\langle{p}\rangle$.
    The dots represent our bootstrap results, while the curves correspond to the Hamiltonian diagonalization results.
We use the same color convention as in Fig. \ref{BB} to indicate the energy bands. }
    \label{kvs}
\end{figure}

\begin{figure}
    \includegraphics[width=0.9\columnwidth]{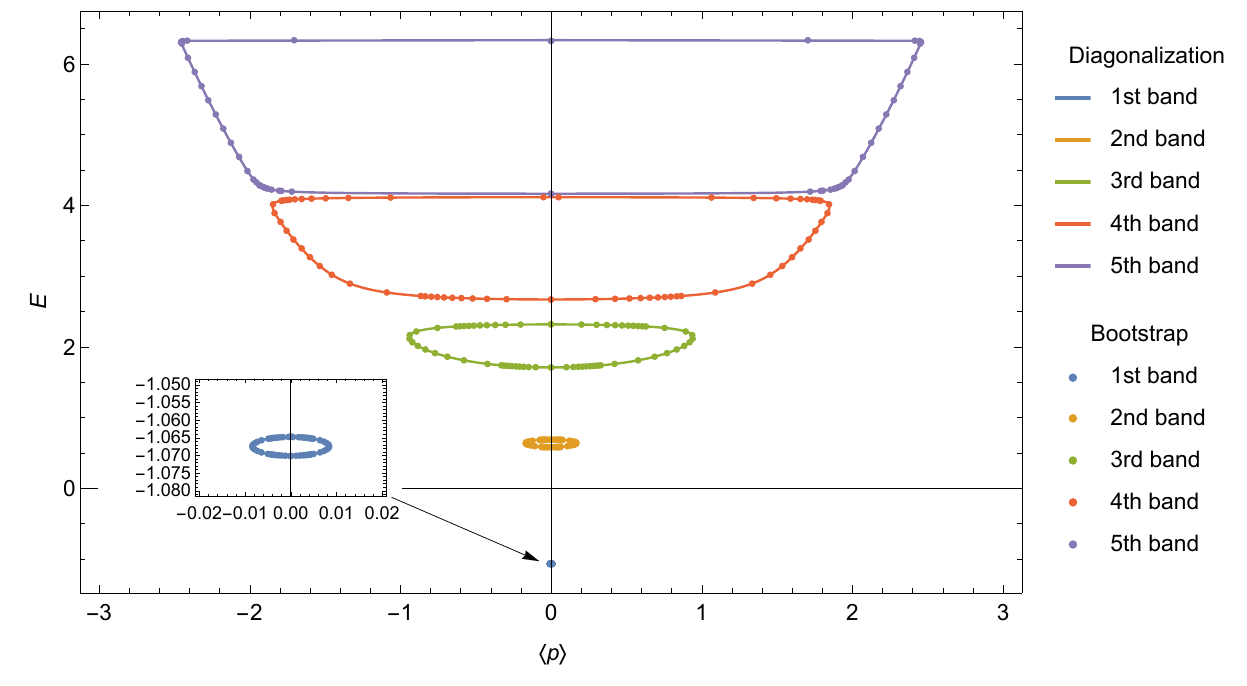}
    \centering
    \caption{Relation between $E$ and $\langle{p}\rangle$.
    The dots are computed by our bootstrap method.
    The curves are deduced from the Hamiltonian diagonalization.
   The  color convention for different energy bands is the same as that in Fig. \ref{BB}. }
    \label{pvs}
\end{figure}

Let us also explore the relationships between other expectation values and $k$. 
The accurate determinations of the Bloch momenta allow us 
to derive the $k$ dependence of $\langle{e^{ix}}\rangle$ and $\langle{p}\rangle$,  
which are presented in Fig. \ref{kvs}. 
We can see that $\langle{e^{ix}}\rangle$ is invariant under the transformation $k\rightarrow -k$, whereas $\langle{p}\rangle$ changes sign. 
As $k$ grows in the positive range, we notice that $\langle{e^{ix}}\rangle$ is monotonically decreasing for the first, third and fifth bands, but increasing for the second and fourth bands. 
In general, $\langle{p}\rangle$ vanishes only at $k=0$ and $|k|=1/2$.
In Fig. \ref{pvs}, we further present the accurate relation between $E$ and $\langle{p}\rangle$. 
Although the Bloch momentum $k$ can take any real value, 
the expectation value of $p$ is restricted to a finite range, which becomes larger for higher energy bands. 
Note that $\langle{p}\rangle$ is related to the dependence of $E$ on the gauge invariant quantity $\langle \dot{x}\rangle$ in 
the $\theta$ term problem \cite{Aikawa:2021eai}. 
In the Fig. 4 of \cite{Aikawa:2021eai}, 
the inside regions were not excluded by the positivity constraints, 
so the positive bootstrap result for $\langle \dot x \rangle$ at a given $E$ becomes a finite range. 
\footnote{The minimum and maximum of an energy band are associated with $\langle{p}\rangle=0$, 
so the positivity bounds with $\langle{p}\rangle=0$ in \cite{Tchoumakov:2021mnh} do not exclude 
the physical range of energy spectra. }
In contrast, our new approach leads to accurate determinations of the relations between $E$ and $\langle{p}\rangle$.

 {
We also compare the computational time of our bootstrap method and the Hamiltonian diagonalization method
for determining the relations among $E,\,k,\,\langle p \rangle,$ and $\langle e^{ix} \rangle$ in the first band.
Our bootstrap method with $A=13$ takes about 10s to obtain the values of $k$, $\langle p \rangle$ and $\langle e^{ix} \rangle$ for given $E$, 
while the Hamiltonian diagonalization method with $D=8$ takes about 0.08s to compute the values of $E$, $\langle p \rangle$ and $\langle e^{ix} \rangle$ for given $k$.
The results are accurate to the same decimal places.
}

\subsection{Fourier series solution}\label{M2}
According to Bloch's theorem, the expectation values should satisfy some periodic constraints
\begin{equation} \label{f-perodic}
e^{-iak}f_{n,a}=\left(e^{-iak}f_{n,a}\right)\Big|_{a\rightarrow a+2\pi}\,,
\end{equation}
so we can solve the differential equations \eqref{pde0} by truncated Fourier series. 
Furthermore, we notice that the combination in \eqref{f-perodic} satisfies
\begin{equation} \label{f-conjugate}
    e^{-iak} f_{n,a}=\left(e^{iak} f_{n,-a}\right)^\ast\,,
\end{equation}
where $\ast$ denotes the complex conjugate 
and we refer to Appendix \ref{ap} for the derivation details. 
We can further use the translation symmetry $f_{n,a+2\pi}=e^{2\pi ik}f_{n,a}$ to derive
\begin{equation}\label{2pi-a}
\left(e^{-iak} f_{n,a}\right )^\ast
= e^{-i(2\pi-a)k} f_{n,2\pi-a}\,,
\end{equation} 
so the case of $a=\pi$ should be a real combination, i.e., $\text{Im}\left(e^{-i\pi k} f_{n,\pi}\right )=0$. 
Accordingly, the truncated Fourier series are 
\begin{equation}\label{Fourex}
    f_{n,a}\equiv \langle e^{inx}e^{iap} \rangle \approx e^{i ak}
     \sum_{m=0}^{F} \left(t_{n,m} \cos m a + i u_{n,m} \sin m a  \right),
\end{equation}
where $t_{n,m}, u_{n,m}$ are real coefficients, and $F$ is the truncation order of the Fourier series. 
For $n\geq 0$, the independent set of $f_{n,a}$ are associated with $n=0,1,2,3$.  

The case of $a=\pi$ is special.   
As mentioned above, we can use \eqref{2pi-a} to deduce that 
$e^{-i \pi k} f_{n,\pi}$ should be real. 
As explained in Sec. \ref{lnepi}, 
we can determine $f_{n,\pi}$ easily using the large $n$ expansion and matching conditions. 
Then we solve the differential equations \eqref{pde0} around $a=\pi$, 
which leads to constraints on the Fourier expansion coefficients $(t_{n,m}, u_{n,m})$.
We choose the set of independent parameters as $(t_{0,0},\,t_{1,0},\,t_{2,0},\, t_{3,0})$, 
which can be expressed in terms of $(f_{0,\pi}, f_{1,\pi}, f_{2,\pi}, f_{3,\pi})$. 
We can further use \eqref{hocpi} to express them in terms of $(E,\,k,\,\lambda_1)$, 
where $\lambda_1$ is a normalization factor. 
To deduce the $(E, k)$ dependence of $f_{n,\pi}$, we choose $N=5$ for the truncation order of the $1/n$ series and $M=11$ for the matching order in \eqref{lneEfpi}. 
In this way, 
we derive the approximate coefficients for the Fourier series in \eqref{Fourex} up to the real parameter $\lambda_1$.

\begin{figure}
    \includegraphics[width=0.8\columnwidth]{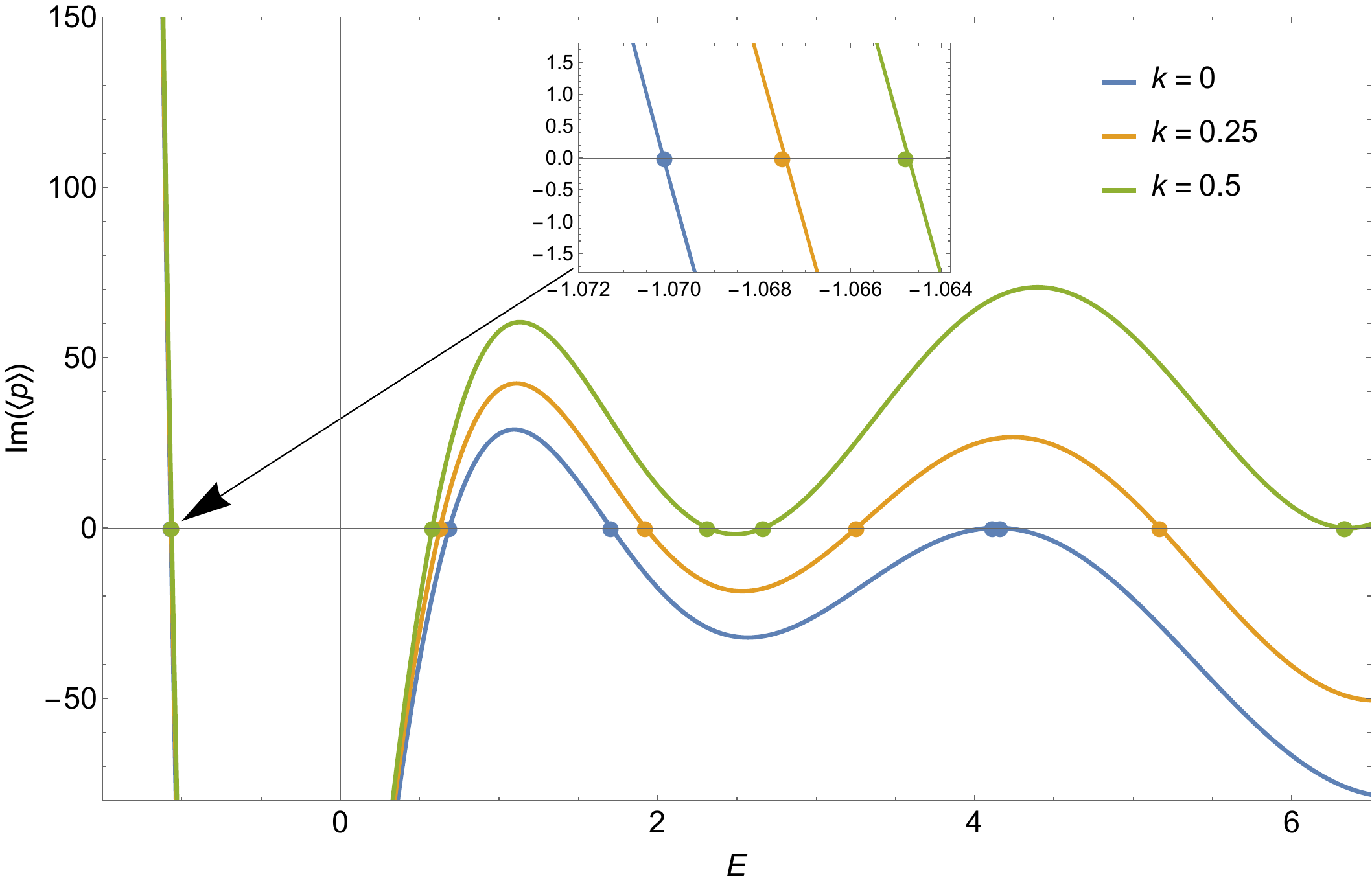}
    \centering
    \caption{The imaginary part of $\langle{p}\rangle$ as a function of $E$. 
    We consider the Bloch momenta $k=0, 0.25,0.5$. 
    The curves are obtained from the bootstrap computation with $N=5,\,M=11,\, A=10,\, F=5$. 
    The intersections with the horizontal line $\text{Im}\langle{p}\rangle=0$ match well with 
    the energies from the Hamiltonian diagonalization (dots).
    }
    \label{f001E}
\end{figure}

\begin{figure}
    \includegraphics[width=0.90\columnwidth]{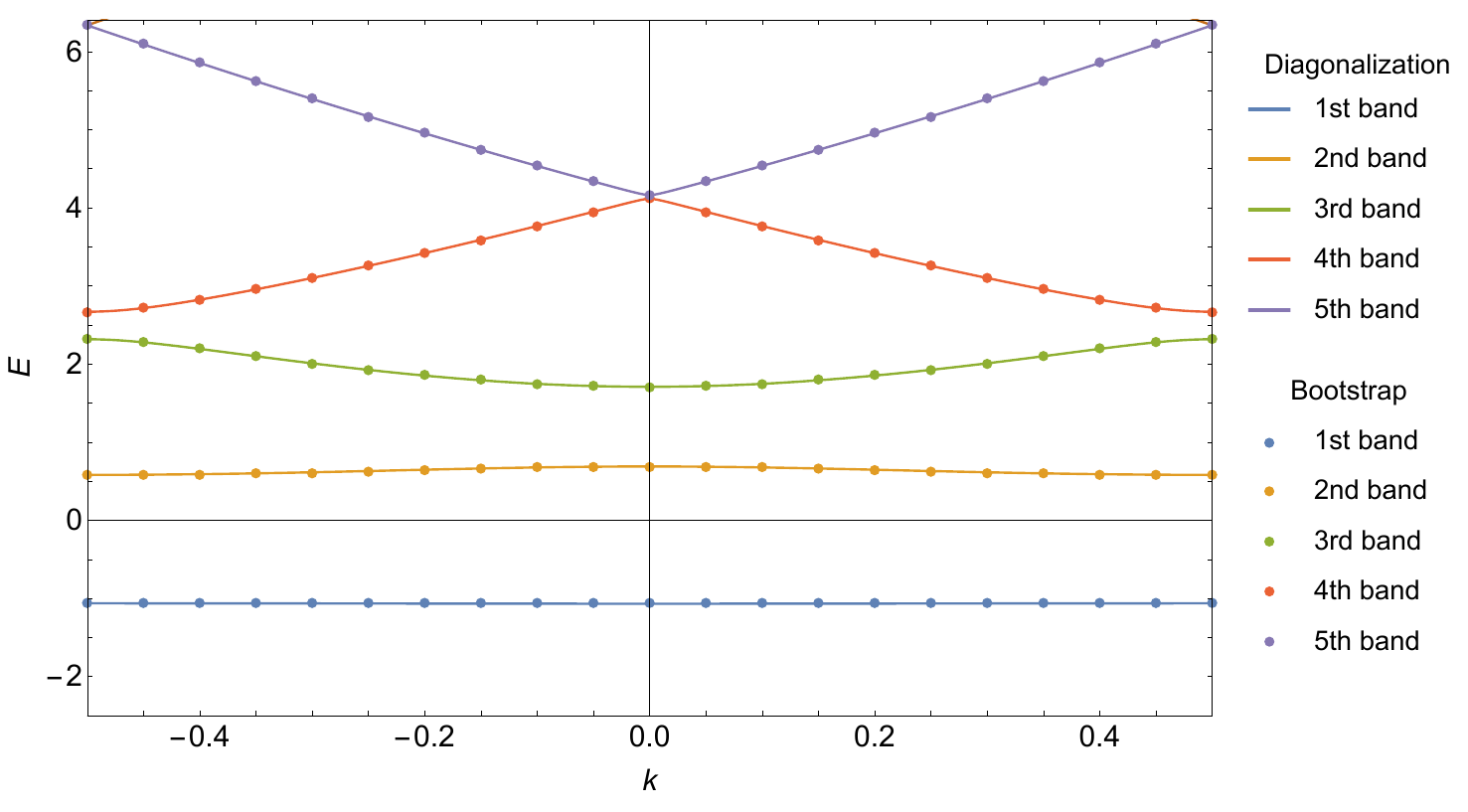}
    \centering
    \caption{$k$ dependence of $E$ from the truncated Fourier series solutions to the differential equations \eqref{pde0} and reality condition $\text{Im}(\langle{p}\rangle)=0$.
    The dots denote our bootstrap results, 
    while the curves indicate the Hamiltonian diagonalization results. 
    We choose an equal spacing for $k$.  
    }
    \label{BB2}
\end{figure}

At this point, the two parameters $(E, k)$ are independent, 
but the dispersion relation indicates that they should be related to each other. 
To determine their relation, 
we further extract $\langle{p}\rangle=f_{0,0,1}$ from the $a=1/10$ results $f_{n,1/10}$  
using the truncated $a$ series in \eqref{smallao}. 
\footnote{Although the truncated Fourier series remain finite in the $a\rightarrow 0$ limit, they cease to be accurate solutions to the differential equations near $a=0$, and the error grows as $a^{-1}$. 
Therefore, we also make use of the small $a$ expansion to connect $\langle{p}\rangle=f_{0,0,1}$ with the $a=1/10$ results.}
As the dispersion relation is independent of the normalization factor $\lambda_1$, 
we set $\lambda_1=1$ for simplicity and impose the correct normalization later. 
Then we consider the reality condition 
\begin{equation}\label{realp}
    \text{Im}(\langle{p}\rangle) = 0\,,
\end{equation} 
which plays the role of a quantization condition for $E$. 
In Fig. \ref{f001E}, we present the imaginary part of $\langle{p}\rangle$ as a function of $E$ for $k=0, 0.25, 0.5$. 
We can see that there are multiple zeros at a given $k$.

The solutions to \eqref{realp} match well with the diagonalization results for $E$. 
In this way, we obtain the accurate relation between $E$ and $k$, 
which is presented in Fig. \ref{BB2}. 
The resulting dispersion relations are again in excellent agreement with the Hamiltonian diagonalization results. 
We further compute the $a$ dependence of $e^{-ika}f_{n,a}$ for the case of $k=0.1$ in the first band
 {and reproduce Fig. \ref{avs},}
which exhibits nice symmetric properties in accordance with \eqref{f-conjugate}, \eqref{2pi-a}. 
We also use the normalization condition \eqref{normalization} to fix $\lambda_1$, 
so $f_{0,0}=e^{-2\pi i k}f_{0,2\pi}=1$.  
The diagonalization results verify explicitly that $e^{-ika}f_{n,a}$ are indeed real at $a=0,\pi,2\pi$. 
 {Furthermore, Fig. \ref{kvs} and Fig. \ref{pvs} can also be reproduced from the truncated Fourier series solutions.
We determine the Fourier coefficients in \eqref{Fourex} 
using \texttt{NSolve} in \textit{Mathematica}.
To achieve the same accuracy, this computation takes approximately the same time as using \texttt{ParametricNDSolve} in Sec. \ref{M1}.}

\section{The Weyl integral and $\langle e^{i\pi p}(ip)^s \rangle$ with noninteger $s$}\label{nonints}
In Sec. \ref{M2}, we use the truncated Fourier series to encode the analytic dependence of $f_{n,a}=\langle e^{inx} e^{iap} \rangle$ on $a$. 
We can introduce $p^s$  by taking derivatives with respect to $a$
\begin{equation}
\frac {\partial^s}{\partial a^s}\langle e^{inx} e^{iap} \rangle=\langle e^{inx} e^{iap} (ip)^s \rangle\,.
\end{equation}
Below, we further extend the domain of $s$ from non-negative integers to complex numbers using fractional calculus. 
For simplicity, we restrict to the case of $k=0$, so $f_{n,a,s}$ is periodic in $a$. 
We confine our discussion to the case of $f_{0,\pi,s}=\langle e^{i\pi p}(ip)^s \rangle$  
for the $k=0$ state in the second energy band. 

For periodic functions, it is natural to make use of the Weyl integral in fractional calculus. 
Suppose that the function $g(a)$ admits a Fourier series expansion
\begin{equation}\label{g-Fourier}
    g(a)=\sum_{m=-\infty}^{\infty} b_m\, e^{i m a}, \quad b_0=0\,.
\end{equation}
The $s$-order Weyl integral of $g(x)$ is defined as
\begin{equation}\label{Weylintegral}
    \frac{d^s}{d a^s}g(a) = \sum_{m=-\infty}^{\infty} (i m)^s b_m\, e^{i m a}\,.
\end{equation}
For positive integer values of $s$, the Weyl integral \eqref{Weylintegral} reduces to the standard $s$th derivative. 
For negative integer values of $s$, 
the Weyl integral \eqref{Weylintegral} can be interpreted as $(-s)$th indefinite integral, which is normalized by integration from $a=0$. 
The absence of an $m=0$ term in \eqref{g-Fourier} is to avoid the divergence issue 
for negative integer values of $s$. 

\begin{figure}
    \centering
    \includegraphics[width=0.6\linewidth]{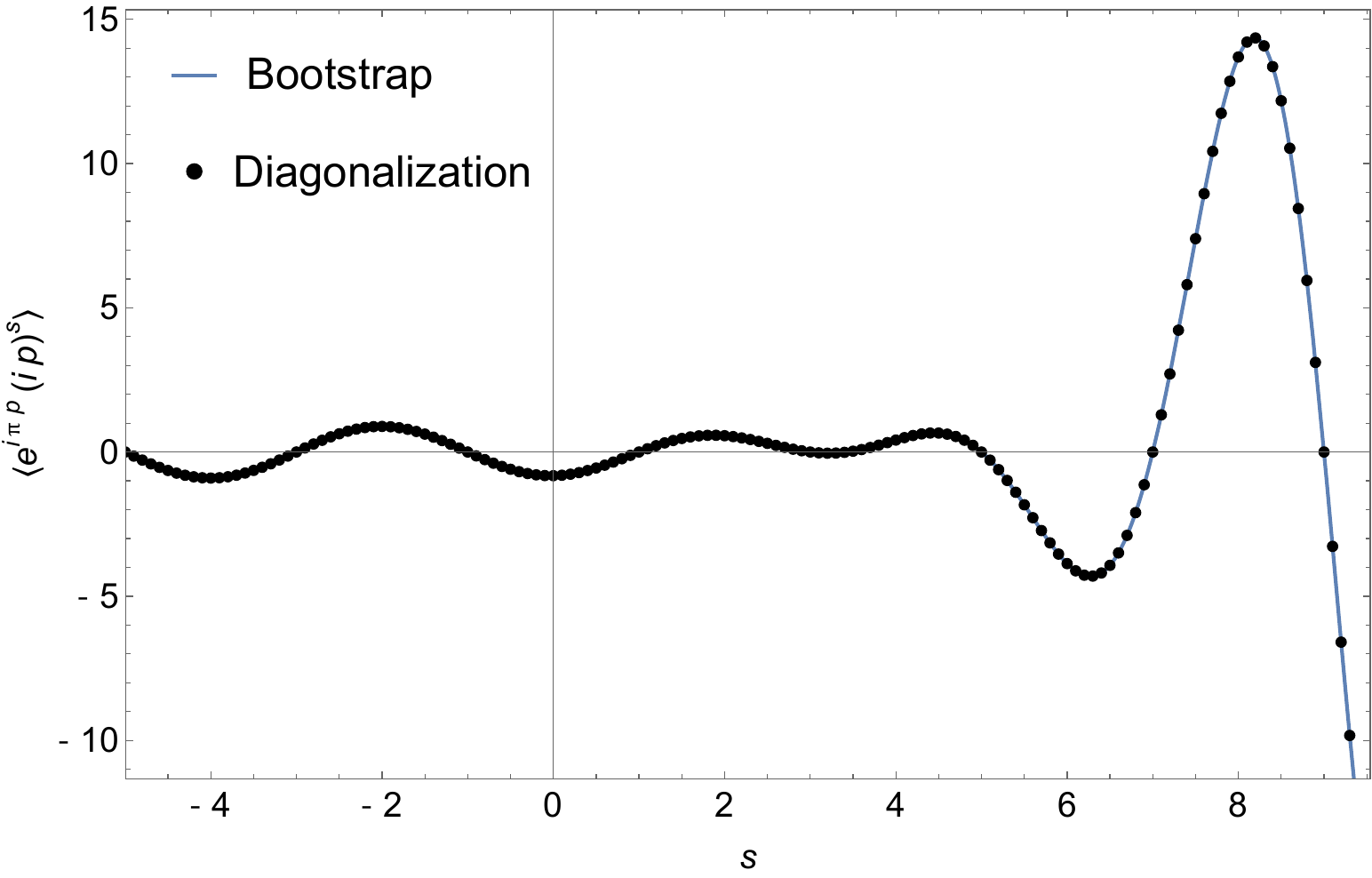}
    \caption{Continous $s$ dependence of $\langle e^{i\pi p}(ip)^s \rangle$ for the $k=0$ state in the second band.
    Note that $\langle e^{i\pi p}(ip)^s \rangle$ is real for real $s$. 
    The blue curve is the bootstrap result, while  
    the black dots denote the Hamiltonian diagonalization results. 
    The curve is obtained from the bootstrap computation with $N=5,\,M=11,\,F=5$. }
    \label{nonintsg}
\end{figure}

\begin{figure}
    \centering
    %\subfloat[The real part]
    \begin{subfigure}{.49\linewidth}
        {\includegraphics[width=0.975\linewidth]{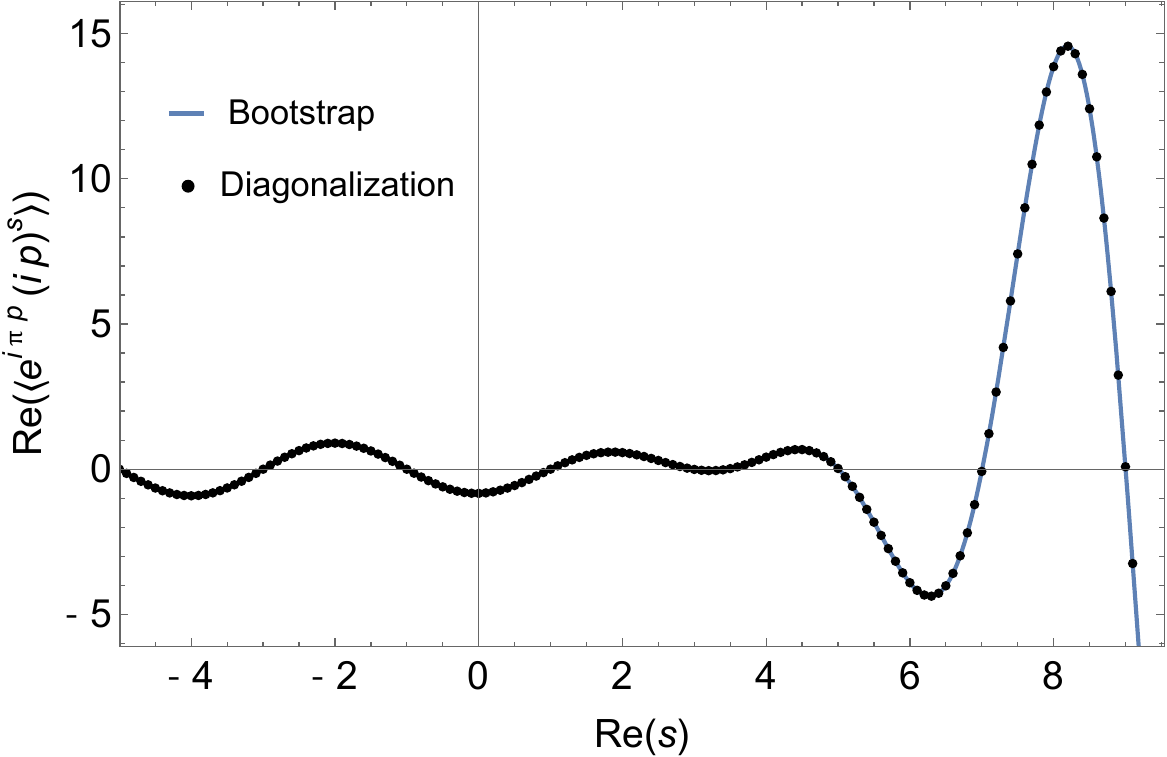}}
        \caption{}
    \end{subfigure}
    %\hspace{0.5cm}
    %\subfloat[The imaginary part]
    \begin{subfigure}{.49\linewidth}
        {\includegraphics[width=0.975\linewidth]{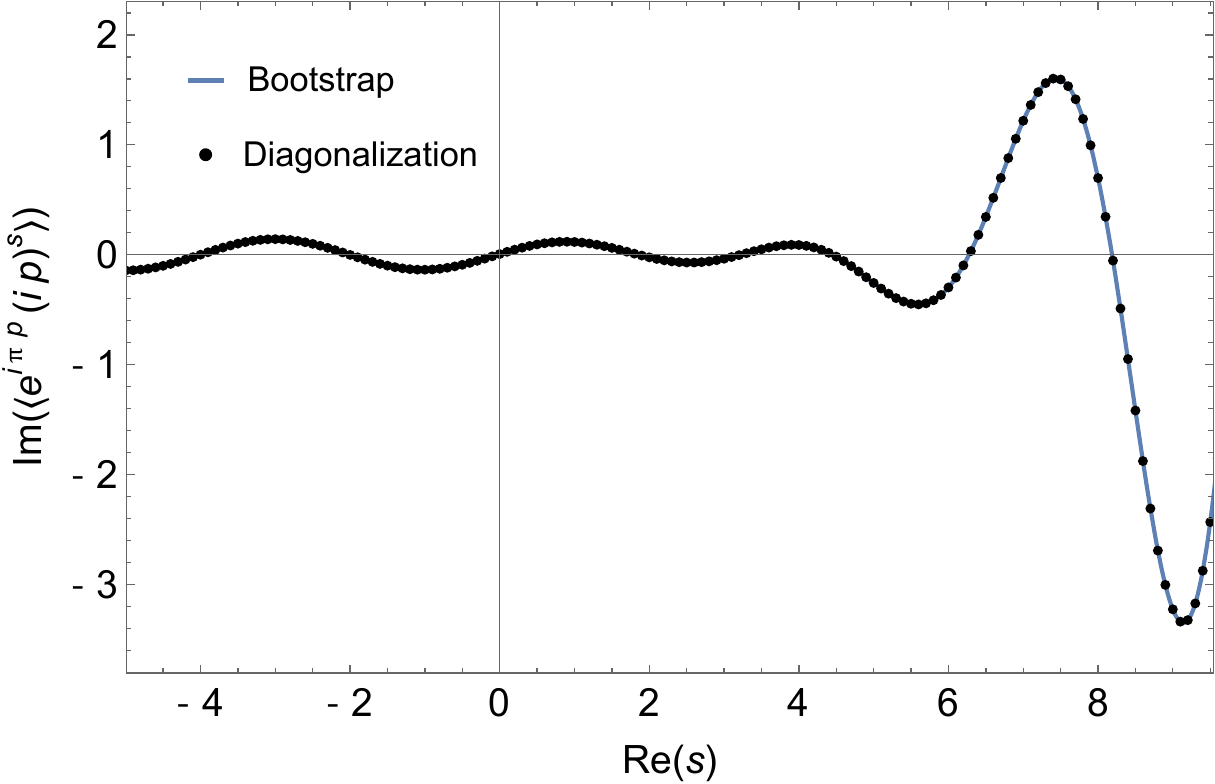}}
        \caption{}
    \end{subfigure}
    \caption{Continous $s$ dependence of $\langle e^{i\pi p}(ip)^s \rangle$ with $s=\text{Re}(s)+i/10$. 
    The blue curves and black dots are associated with the bootstrap computation and the Hamiltonian diagonalization, respectively. 
    (a) The real part.
    (b) The imaginary part.}
    \label{Ims}
\end{figure} 

It is straightforward to obtain $\langle e^{i\pi p}(ip)^s \rangle$ with noninteger $s$ using \eqref{Fourex}. We just need to express the trigonometric functions  in terms of exponential functions, and then use the Weyl integral formula \eqref{Weylintegral}. 
For comparison, we also consider the noninteger $s$ extension in the Hamiltonian diagonalization approach \eqref{diagH}.
In this case, a power term $(ip)^s=\frac {d^s}{dx^s}$ can be interpreted as an $s$th derivative acting on the wave function. 
We consider the $k=0$ state in the second band because $b_0=0$
and the Weyl integral formula \eqref{Weylintegral} is applicable. 

In Fig. \ref{nonintsg}, we present the results of $\langle e^{i\pi p}(ip)^s \rangle$ for real $s$,
where we consider the $k=0$ state in the second band. 
According to \eqref{Weylintegral}, 
the rapidly oscillatory modes become more and more important at large $s$, 
while the $s\to -\infty$ limit is dominated by the $e^{\pm ix}$ terms.
We also notice that $f_{0,\pi,s}$ vanishes when $s$ is odd.  
As the Fourier series \eqref{Fourex} contains only the cosine terms, 
the Weyl integral vanishes at odd integer $s$ for $a=\pi$. 
Furthermore, we can consider the situation where $s$ is a complex number.  
In Fig. \ref{Ims}, we present the results for the case of complex $s=\text{Re}(s)+\frac i {10}$, 
where the imaginary part of $s$ is set to $1/10$. 
For complex $s$, the imaginary part of $\langle e^{i\pi p}(ip)^s \rangle$ does not vanish automatically, 
so the real part and imaginary part are presented separately.  
In contrast to the real $s$ case, the real part of $\langle e^{i\pi p}(ip)^{\text{Re}(s)+\frac i{10}} \rangle$ does not vanish at odd integer $\text{Re}(s)$. 
In both Fig. \ref{nonintsg} and Fig. \ref{Ims}, 
the bootstrap results are in good agreement with the diagonalization results. 
Again, we can readily obtain the continuous $s$ dependence in the bootstrap approach, 
while the diagonalization approach requires evaluating different numerical integrals at different $s$. 

\section{Discussion}\label{conclusion}
In summary, we developed a new bootstrap procedure for periodic quantum mechanical systems. 
A novel element in the bootstrap formulation is an enlarged set of operators $\{e^{inx}e^{iap}p^s\}$, which includes the translation operator $e^{iap}$.  
Their expectation values satisfy self-consistency constraints in the form of recursion relations in $n$ and differential equations in $a$. 
In some limits, they can be solved analytically by the large $n$ expansion or small $a$ expansion. 
At some fixed $a$, we used the matching conditions at finite $n$ to determine the free parameters, 
which provide boundary conditions for the differential equations. 
Without resorting to explicit wave functions, 
we successfully extract the Bloch momentum $k$ from the $a$ dependence of $\langle e^{inx}e^{iap} \rangle$ and the reality conditions on $(E, k, \langle{p}\rangle)$. 
This procedure also applies to the closely related problems for a quantum particle on a circle and a quantum mechanical analog of the $\theta$ term. 
We obtain the accurate dispersion relations and the $k$ dependence of other expectation values, 
resolving the problems raised in \cite{Tchoumakov:2021mnh,Berenstein:2021loy, Aikawa:2021eai}. 
Furthermore, we considered the noninteger power of the momentum operator, i.e., $(ip)^s$ with noninteger $s$, using the Weyl integral in fractional calculus. 
 
A natural extension is to consider quasiperiodic problems, i.e., incommensurate systems.  
It is interesting to see if the bootstrap approach can capture 
Hofstadter's butterfly \cite{PhysRevB.14.2239}, the localization transition 
\cite{Abal,Harper:1955jqr},  
and multifractality \cite{doi:10.1098/rspa.1984.0016}. 
Another curious question is how to understand the geometric and topological aspects of the band structure from the bootstrap perspective, 
such as the Berry connection and the Berry phase,  which are usually extracted from wave functions. 
The Su-Schrieffer-Heeger model \cite{Su:1979ua} may provide a simple playground for bootstrapping topological phases and their transitions. 

A more ambitious direction is to extend the one-body periodic bootstrap method to the cases of quantum many-body systems and quantum field theories, such as the strongly correlated electron systems 
and the $\theta$ term in quantum chromodynamics. 
See \cite{han2020quantummanybodybootstrap} and \cite{Gao:2024etm} for 
bootstrap studies of the Hubbard model and quantum Hall systems. 
The nonperturbative bootstrap investigations of lattice gauge theories \cite{Anderson:2016rcw,Kazakov:2022xuh,Kazakov:2024ool,Li:2024wrd,Guo:2025fii} 
and  lattice spin models \cite{Cho:2022lcj,Nancarrow:2022wdr,Berenstein:2024ebf} are also of significant importance. 
Another related direction of interest is to study frustrated systems \cite{TOULOUSEG1977,JVillain_1977} from the bootstrap perspective.  
It may be useful to develop a more sophisticated implementation of translation operations along the lines of the present work. 

\section*{Acknowledgments}
This work was supported by 
the Natural Science Foundation of China (Grants No. 12522504 and No. 12205386) and 
the Guangzhou Municipal Science and Technology Project (Grant No. 2023A04J0006).

\appendix

\section{An identity for $\langle e^{inx} e^{ia(p-k)} \rangle$}\label{ap}
Let us prove the identity
\begin{equation}\label{ap-identity}
    e^{-iak} f_{n,a}=(e^{iak} f_{n,-a})^*.
\end{equation}
According to Bloch's theorem \eqref{blochth}, we have
\begin{equation}\label{eiakf-bloch}
        e^{-iak} f_{n,a} =e^{-iak} \int_{0}^{2\pi} \psi^*(x) e^{inx} \psi(x+a) dx
        =\int_{0}^{2\pi} e^{inx} \phi_k^*(x)\phi_k(x+a) dx\,.
\end{equation}
The complex conjugate of \eqref{eiakf-bloch} gives
\begin{align}
\left(e^{-iak} f_{n,a}\right )^\ast=&\int_{0}^{2\pi} e^{-inx} \phi_k(x)\phi_k^*(x+a) dx
\nonumber\\
=&\int_{-2\pi}^{0} e^{inx} \phi_k(-x)\phi_k^*(-x+a) dx
\nonumber\\
=&\int_{0}^{2\pi} e^{inx} \phi_k(-x)\phi_k^*(-x+a) dx\,,
\end{align} 
where we used $e^{in(x+2\pi)}=e^{inx}$ and $\phi(x+2\pi)=\phi(x)$. 
As the Hamiltonian in the basis $\left\{\frac{1}{\sqrt{2\pi}}e^{ikx}e^{i m x}\right\}$ is a real and symmetric matrix,  
we can impose that the eigenfunctions 
\begin{equation}
    \phi_k(x)=\sum_{m} c_{k,m} e^{i m x}
\end{equation}
have real coefficients $c_{k,m}$, so $\phi_k(-x)=\phi_k^\ast(x)$. 
We have
\begin{equation}
\left(e^{-iak} f_{n,a}\right )^\ast
=\int_{0}^{2\pi} e^{inx} \phi_k^\ast(x)\phi_k(x-a) dx
=e^{iak} f_{n,-a}\,.
\end{equation}

\end{document}